\documentclass[aps,prd,nofootinbib,showpacs,preprintnumbers,amsmath,amssymb]{revtex4-1}
\usepackage{epsfig}
\usepackage{graphics}
\usepackage{graphicx}
\usepackage{amssymb}
\usepackage{latexsym}
\usepackage{multirow}
\usepackage{graphicx,color}
\def\be{\begin{equation}}
\def\ee{\end{equation}}
\def\bea{\begin{eqnarray}}
\def\eea{\end{eqnarray}}
\def\bes{\begin{subequations}}
\def\ees{\end{subequations}}
\def\gsim{~\rlap{$>$}{\lower 1.0ex\hbox{$\sim$}}\;}

\begin{document}

\title{CP violation in the rare Higgs decays via exchange of on-shell almost degenerate Majorana neutrinos, $H \to \nu_k N_j \to \nu_k \ell^{-} U {\bar D}$ and $H \to \nu_k N_j \to \nu_k \ell^{+} {\bar U} D$}

\author{Gorazd Cveti\v{c}$^a$}
 \email{gorazd.cvetic@usm.cl}
\author{ C.~S.~Kim$^b$}
\email{ cskim@yonsei.ac.kr}
\author{Jiberto Zamora-Sa\'a$^{c,d}$}
\email{jilberto.zamora@unab.cl}

\affiliation{$^a$ Department of Physics, Universidad T{\'e}cnica Federico Santa Mar{\'\i}a,  Casilla 110-V, Valpara{\'\i}so, Chile\\
$^b$  Department of Physics and IPAP, Yonsei University, Seoul 120-749, Korea\\
$^c$ Center for Theoretical and Experimental Particle Physics (CTEPP) and Department of Physics, Universidad Andres Bello, Fernandez Concha 700, Santiago, Chile\\
$^d$ Millennium Institute for Subatomic physics at high
energy frontier - SAPHIR, Fernandez Concha 700, Santiago, Chile.
}

\begin{abstract}
  We investigate rare decays of Higgs via exchange of two almost degenerate heavy on-shell Majorana neutrinos $N_j$ ($j=1,2$): $\Gamma_{\pm} = \Gamma(h \to \nu_k N_j \to \nu_k \ell^{\pm} \pi^{\mp})$, and into the open quark channels $\Gamma_{\pm} = \Gamma(h \to \nu_k N_j \to \nu_k \ell^{\pm} U D)$, where $U D$ are two jets of open quarks (${\bar U} D$, or $U {\bar D}$, where $U=u, c$ and $D=d, s$). The related CP violation asymmetry $A_{\rm CP} = (\Gamma_{-} - \Gamma_{+})/(\Gamma_{-} + \Gamma_{+})$ is studied in detail. We take into account the $N_1$-$N_2$ overlap and oscillation effects. We can see that for certain, presently acceptable, range of input parameters, such decays with open quark channels, and their asymmetries, could be detected in the International Linear Collider (ILC).
\end{abstract}
\maketitle

\section{Introduction}
\label{sec:Intro}

The Standard Model (SM) of particle physics is the most successful theory of modern particle physics. The SM has been capable to provide high-precision predictions which are in agreement with decades of experimental data. The particles' content of the SM was complete in 2012 by the discovery of the Higgs boson at the Large Hadron Collider \cite{ATLAS:2012yve,CMS:2012qbp}. However, despite the remarkable success of this model, there is still experimental evidence that can not be accommodated with it. Among such evidence are Neutrino Oscillations (NOs), Dark Matter (DM) and Baryonic Asymmetry of the Universe (BAU).

In the last years, neutrino oscillations experiments have demonstrated that
active neutrinos ($\nu$) are very light massive particles ($m_{\nu} < 1$ eV)~\cite{Fukuda:1998mi,Eguchi:2002dm} as opposed to the SM where they are massless; therefore, the SM must be extended. Among the most promising extensions to the SM, which explain very light massive neutrinos, are the models with the See-Saw Mechanism (SSM) \cite{Mohapatra:2005wg,Mohapatra:2006gs}. The SSM works by introducing a new Majorana neutral lepton (SM-singlet) called heavy sterile neutrino (HSN), with a highly suppressed interaction with gauge bosons ($Z,W^{\pm}$) and the other leptons ($e, \mu, \tau$). While the Dirac neutrinos can only participate in processes that conserve the lepton number (LNC), the Majorana neutrinos can induce both lepton number conserving and lepton number violating (LNV) processes, which opens a plethora of new physics. Despite the suppression mentioned above, HSNs can be searched at colliders~\cite{Milanes:2016rzr,Tapia:2021gne,Das:2018usr,Das:2017nvm,Das:2012ze,Antusch:2017ebe,Das:2017rsu,
Das:2017zjc,Chakraborty:2018khw,Cvetic:2019shl,Antusch:2016ejd,Cottin:2018nms,Duarte:2018kiv,Drewes:2019fou,
Cvetic:2018elt,Cvetic:2019rms,Das:2018hph,Das:2016hof,Das:2017hmg}, rare meson decays~\cite{CPVBelle,Dib:2000wm,Cvetic:2012hd,GCCSKJZS1,GCCSKJZS2,symm,oscGCetal,Moreno:2016cfz,Milanes:2018aku,Mejia-Guisao:2017gqp} and tau factories \cite{Zamora-Saa:2016ito,Kim:2017pra,Tapia:2019coy,Dib:2019tuj}.

One of the most well reputed neutrino mass models, based on SSM, is the Neutrino-Minimal-Standard-Model, $\nu$MSM~\cite{Asaka:2005an,Asaka:2005pn}, which by introducing two almost degenerate HSNs, that can oscillate among themselves, and with masses $M_{N1} \approx M_{N2} \sim 1 $ GeV, could lead to a successful BAU via leptogenesis \cite{Akhmedov:1998qx} and also provide a natural DM candidate by adding a third HSN with mass $M_{N3} \sim 1$ keV. In addition, according to Sakharov conditions \cite{Sakharov:1967dj}, CP invariance must be broken in order to produce a successful baryon asymmetry of the universe. In last years, NOs experiments have indicated that mixing-angle $\theta_{13}$ is nonzero~\cite{An:2012eh}, suggesting the possibility of CP violation in the light neutrino sector \cite{Abe:2018wpn}. However, this light neutrino sector CP-violation source is not enough, and other sources of CP violation are required in order to explain the BAU via leptogenesis (see \cite{Chun:2017spz} for a review). Furthermore, when the HSN masses are below the electroweak scale ($M_N < 246$ GeV), the BAU could be generated via CP-violating Heavy Neutrino Oscillations (HNOs) \cite{Akhmedov:1998qx,Drewes:2016gmt}. In the present work, we will consider this range of HSN masses and thus, this particular scenario of CP violation.

In Ref.~\cite{Das:2017zjc} the HSM production in Higgs decays has been studied, mediated by  the Dirac Yukawa coupling in the singlet seesaw extension of the SM, considering the LHC Higgs data at $\sqrt{s}=8$ TeV. That work has presented new limits for active-sterile neutrino mixings, particularly for $e$ and $\mu$ flavours for $10 \ {\rm GeV} \leq M_N \leq 120 \ {\rm GeV}$. They also have shown that more than $3 \sigma$ significance could be reached for HSN mass ranging between 70 and 120 GeV. However, in this work, we shall pay our attention to CP violation in rare Higgs decays for HSN masses below 80 GeV, considering the expected numbers of produced Higgs bosons at ILC (see Table I in Sec.~\ref{sec:ANARES}).

\begingroup \color{black} As is well known~\cite{Gunion:1989we}, Higgs decays in the minimal Higgs model within the SM
cannot induce CP violation. We will show a new mechanism of CP violation in Higgs decays even without extending 
the Higgs potential (which includes the CP-violating coupling constant) and/or without extending Higgs structure to 
two or more doublets (which include  CP-even and CP-odd Higgs). We will show the new CP violation of Higgs decays which
comes through the on-shell Majorana HSN decays.
\endgroup

The work is organised as follows. In Sec.~\ref{sec:Gnoosc}, we study the HSN decays without oscillation effects. In Sec.~\ref{sec:Gosc}, we include the oscillation effects. In Sec.~\ref{sec:ANARES}, we present the analysis and results, and in Sec.~\ref{sec:concl} we summarise our conclusions.

\section{The general formula for  $\Gamma_{\pm} = \Gamma(h \to \nu_k N_j \to \nu_k \ell^{\pm} \pi^{\mp})$, and for $\Gamma_{\pm} = \Gamma(h \to \nu_k N_j \to \nu_k \ell^{\pm} U D)$, without the oscillation effects}
\label{sec:Gnoosc}

The Yukawa interaction of Higgs doublet $\Phi^T = (\phi_{+}, \phi_0)$ in SM with the $\ell$-generation of leptons ($\ell=e, \mu, \tau$) is
\be
{\cal L}^{(\ell)}_{Y,SM} = - G_{\ell}  \left[ {\bar \ell_R} ( \Phi^{\dagger} L) + ({\bar L} \Phi) \ell_R \right],
\label{LYukSM1a} \ee
  where $L^T = (\nu_{\ell}, \ell)_{L}$ and $R=\ell_R$, and the flavour-neutrino $\nu_{\ell}$ has no right-handed component. In the unitary gauge, $\Phi^T \to  (1/\sqrt{2}) (0, v + h)$ where $v \approx 246$ GeV is VEV and $h$ is the physical SM Higgs,  and the Yukawa terms become
\be
{\cal L}^{(\ell)}_{Y,SM}  = -   \frac{G_{\ell}}{\sqrt{2}} (v + h) ({\bar \ell}_R \ell_L + {\bar \ell}_L \ell_R) = - \frac{G_{\ell} v}{\sqrt{2}} {\bar \ell} \ell -\frac{G_{\ell}}{\sqrt{2}} h {\bar \ell} \ell.
 \label{LYukSM1b} \ee
 Here, the mass of the charged lepton is $m_{\ell} = G_{\ell} v/\sqrt{2}$.

 On the other hand, when we also have right-handed light neutrinos $\nu_{\ell,R}$ ($R=\ell_R$ and $R=\nu_{\ell,R}$ are both SM singlets), we have, in addition to the Yukawa terms Eq.~(\ref{LYukSM1a}), also the following Yukawa terms which involve the Higgs doublet complex conjugate ${\widetilde \Phi}^T= (\phi_0^{\ast}, -\phi_{+}^{\ast}) =  (\phi_0^{\ast}, -\phi_{-})$:
\be
{\cal L}^{(\nu_{\ell})}_{Y} = - G_{\nu_{\ell}}  \left[ {\bar \nu_{\ell,R}} ( {\widetilde \Phi}^{\dagger} L) + ({\bar L} {\widetilde \Phi}) \nu_{\ell,R}  \right],
\label{LYukSM2a} \ee
which in the unitary gauge acquires the form analogous to Eq.~(\ref{LYukSM1b})
\be
{\cal L}^{(\nu_{\ell})}_{Y}  = -   \frac{G_{\nu_{\ell}}}{\sqrt{2}} (v + h) ({\bar \nu_{\ell,R}} \nu_{\ell,L} + {\bar \nu_{\ell,L}} \nu_{\ell,R}) = - \frac{G_{\nu_{\ell}} v}{\sqrt{2}} {\bar \nu_{\ell}} \nu_{\ell} -\frac{G_{\nu_{\ell}}}{\sqrt{2}} h {\bar \nu_{\ell}} \nu_{\ell}.
\label{LYukSM2b} \ee
The (real) couplings $G_{\nu_{\ell}}$ are in principle unknown, and are not necessarily directly related with the masses of light neutrinos $\nu_k$ ($k=1,2,3$). In our work we will assume that the neutrinos are Majorana. In such a case, the masses $m_{\nu_k}$ of light neutrinos are not equal to the Dirac masses $m_{D,k} = G_{\nu\ell} v/\sqrt{2}$ appearing in Eq.~(\ref{LYukSM2b}), but result from a seesaw mechanism where these Dirac masses are just an element of the mechanism. Nonetheless, a reasonable assumption could be made, namely that $G_{\nu \ell} \approx G_{\ell}$, which suggests that $G_{\nu \tau}$ is the dominant among the three couplings  $G_{\nu \ell}$, i.e.,
\be
G_{\nu_{\tau}} \sim G_{\tau}  \sim \frac{m_{\tau} \sqrt{2}}{v}  \; (\approx 0.01).
\label{Gnutauest} \ee
We do not refer to any specific model for the Yukawa couplings $G_{\nu_{\tau}}$, but will consider that the estimate Eq.~(\ref{Gnutauest}) is valid; specifically, we will consider scenarios with $G_{\nu_{\tau}} \approx 0.01$-$0.03$. We will consider a scenario where we have, in addition to the three light mass eigenstate Majorana neutrinos $\nu_k$ ($k=1,2,3$), at least two additional heavy Majorana neutrinos $N_j$ with masses $M_j \sim 10^1$ GeV ($j=1,2$). In such scenarios, the three flavour eigenstate neutrinos $\nu_{\ell}$ ($\ell=e, \mu, \tau$) have small admixture of the mentioned heavy mass eigenstates
\be
\nu_{\ell} = \sum_{k=1}^3 U_{\ell \nu_k} \nu_k + \sum_{j=1}^2 U_{\ell N_j} N_j,
\label{mix} \ee
where the heavy-light mixing coefficients $U_{\ell N_j}$ are very small. The mixing (\ref{mix}) then implies that the Yukawa coupling in Eq.~(\ref{LYukSM2b}) results in nonzero $U_{\ell N_j}$-suppressed coupling of Higgs $h$ field to the heavy Majorana neutrinos $N_j$
\bes
\label{YuN}
\bea
    {\cal L}^{(h \nu N)}  &=& -\frac{1}{\sqrt{2}}  \sum_{\ell=e,\nu,\tau} G_{\nu_{\ell}} \sum_{j=1}^2 \sum_{k=1}^3 h \left( U_{\ell N_j} U^{\ast}_{\ell \nu_k} {\bar \nu}_k N_j +
 U^{\ast}_{\ell N_j} U_{\ell \nu_k} {\bar N}_j \nu_k    \right)
\label{YuNa} \\
& = & -\sqrt{2} \sum_{j=1}^2 \sum_{k=1}^3 {\cal A}_j^{(k)} \; h {\bar \nu}_k N_j,
\label{YuNb} \eea \ees
where
\be
{\cal A}_j^{(k)} = \sum_{\ell=e,\mu,\tau} G_{\nu_{\ell}} {\rm Re} \left( U_{\ell N_j} U^{\ast}_{\ell \nu_k} \right) \qquad (j=1,2; \; k=1,2,3).
\label{Aj} \ee
The simpler form (\ref{YuNb}) [with (\ref{Aj})] is obtained from the form (\ref{YuNa}) because ${\bar N}_j \nu_{k} \equiv {\overline {\nu^c_{k}}} N_j^c$ and, as the neutrinos are considered to be Majorana, we have $N_j^c=N_j$ and $\nu^c_k= \nu_k$. For definiteness, we could consider that $G_{\nu_{\tau}}$ is the dominant coupling, i.e.,
\bea
{\cal L}^{(h \nu N)}  & \approx & - \sqrt{2} G_{\nu_{\tau}} \sum_{j=1}^2 \sum_{k=1}^3 {\rm Re}( U_{\tau N_j} U^{\ast}_{\tau \nu_k} )\; h {\bar \nu}_k N_j.
\label{YuNtau} \eea
However, our formulas do no depend explicitly on such a choice, and we will use in our formulas the sum, Eqs.~(\ref{YuNb}) and (\ref{Aj}). We recall that ${\cal A}^{(k)}_j$ is a sum of real numbers.

\begin{figure}[htb] 
\centering\includegraphics[width=100mm]{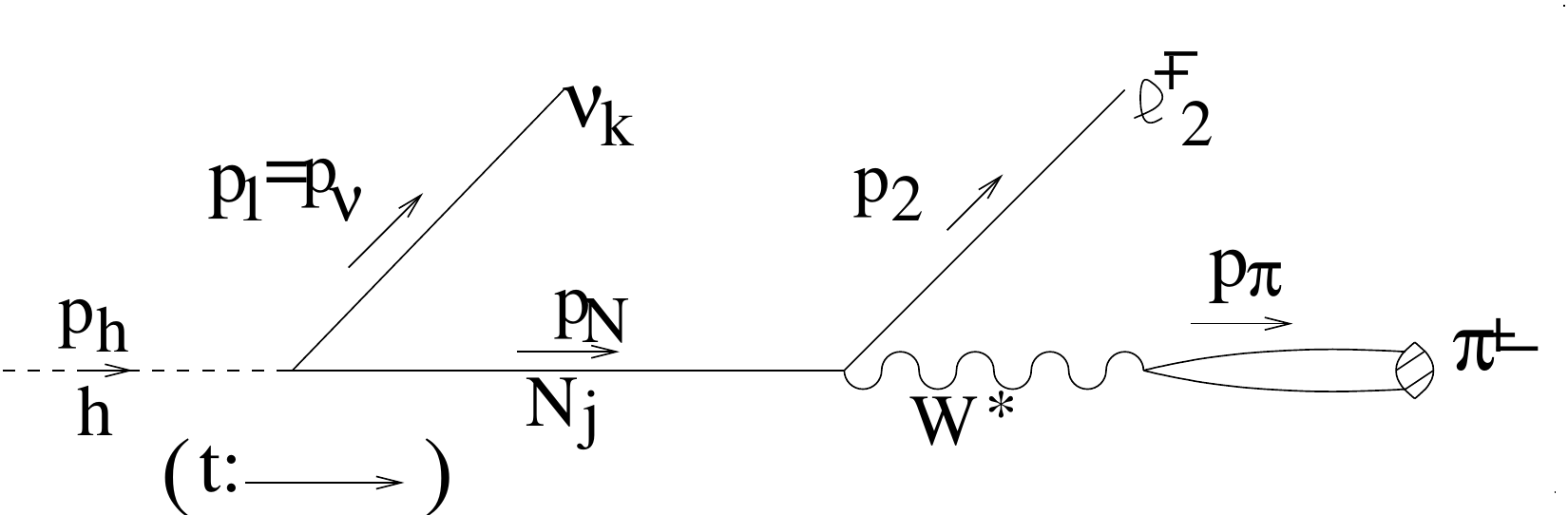}
\caption{\footnotesize The decay  $h(p_h) \to \nu_k(p_1) N_j(p_N) \to \nu_k(p_1) \ell_2^{\mp}(p_2) \pi^{\pm}(p_{\pi})$, where $N_j$ is considered on-shell (i.e., $M_{N_j} < M_h$).}
\label{Fighdec}
\end{figure}
We will neglect the mass of the light neutrino $\nu_k$ produced in the considered decay.
The ${\cal T}_{-}$ matrix element for the decays $h(p_h) \to \nu_k(p_1) N_j(p_N) \to \nu_k(p_1) \ell_2^{-}(p_2) \pi^{+}(p_{\pi})$, cf.~Fig.~\ref{Fighdec}, is\footnote{We use the convention that the ${\cal T}$ and the $S$ decay matrix are related via: $S= i (2 \pi)^4 \delta(P_f-P_i) {\cal T}$.}
\bea
{\cal T}_{-} & = & i K \sum_{j=1}^2 U_{\ell_2 N_j} {\cal A}^{(k)}_j \frac{ \left( 1 - \frac{m_{\pi}^2}{M_W^2} \right) }{ \left( 1 -  \frac{m_{\pi}^2}{M_W^2} - i \frac{\Gamma_W}{M_W} \right) }
    \left[ {\bar u}_{\nu_k}(p_1) ({\displaystyle{\not}p}_h - {\displaystyle{\not}p}_1+M_{N_j}) {\displaystyle{\not}p}_{\pi} (1 + \gamma_5) v_{\ell_2}(p_2) \right] P_j((p_h-p_1)^2),
    \nonumber\\
  \label{Tm} \eea
where
\be
K \equiv G_F f_{\pi} V_{u d},
\label{Kdef} \ee
$G_F=1.1664 \times 10^{-5} \ {\rm GeV}^{-2}$ is the Fermi coupling constant, $f_{\pi}=0.1304$ GeV is the pion decay constant, and $P_j(p_N^2)$ is the denominator of the $N_j$ propagator
\be
P_j(p_N^2) \equiv \frac{1}{( p_N^2 - M_{N_j}^2 + i \Gamma_{N_j} M_{N_j})} \equiv P_j.
\label{Pj} \ee
The  ${\cal T}_{+}$ matrix element for the charge-conjugate decays $h(p_h) \to \nu_k(p_1) N_j(p_N) \to \nu_k(p_1) \ell_2^{+}(p_2) \pi^{-}(p_{\pi})$ is
\bea
{\cal T}_{+} & = & -i K \sum_{j=1}^2 (U_{\ell_2 N_j})^{\ast} {\cal A}^{(k)}_j \frac{ \left( 1 - \frac{m_{\pi}^2}{M_W^2} \right)}{\left( 1 -  \frac{m_{\pi}^2}{M_W^2} - i \frac{\Gamma_W}{M_W} \right)}
\left[ {\bar u}_{\ell_2}(p_2) {\displaystyle{\not}p}_{\pi} (1 + \gamma_5) ({\displaystyle{\not}p}_{h} - {\displaystyle{\not}p}_1+M_{N_j}) v_{\nu_k}(p_1) \right] P_j((p_h-p_1)^2),
\nonumber\\
\label{Tp} \eea
where we note that $ {\cal A}^{(k)}_j$ is a real number, cf.~Eq.~(\ref{Aj}).

When squaring and integrating over the final space phase, we will obtain the corresponding decay width.

When we first square the amplitudes (\ref{Tm}) and (\ref{Tp}) and sum over the helicities of the final state leptons, we obtain
\bes
\label{Tsq}
\bea
| {\cal T}_{-}|^2 & = & K^2 \frac{\left( 1 - \frac{m_{\pi}^2}{M_W^2} \right)^2 }{ \left[ \left( 1 -  \frac{m_{\pi}^2}{M_W^2} \right)^2 + \left( \frac{\Gamma_W}{M_W} \right)^2 \right]} \sum_{i,j=1}^2 U_{\ell_2 N_j}  U_{\ell_2 N_i}^{\ast} {\cal A}^{(k)}_j {\cal A}^{(k)}_i  {\bar t}_{j i} P_j P_i^{\ast}
\label{Tmsq}
\\
| {\cal T}_{+}|^2 & = & K^2 \frac{\left( 1 - \frac{m_{\pi}^2}{M_W^2} \right)^2 }{ \left[ \left( 1 -  \frac{m_{\pi}^2}{M_W^2} \right)^2 + \left( \frac{\Gamma_W}{M_W} \right)^2 \right]} \sum_{i,j=1}^2 U_{\ell_2 N_j}^{\ast}  U_{\ell_2 N_i} {\cal A}^{(k)}_j {\cal A}^{(k)}_i  {\bar t}_{i j} P_j P_i^{\ast} ,
\label{Tpsq}
\eea \ees
where ${\bar t}_{ij}$ are the traces
\bes \label{bart}
\bea
{\bar t}_{ij} & = & {\rm tr} \left[ {\displaystyle{\not}p}_1 ({\displaystyle{\not}p}_{h} - {\displaystyle{\not}p}_1+M_{N_i}) {\displaystyle{\not}p}_{\pi} {\displaystyle{\not}p}_2 {\displaystyle{\not}p}_{\pi} (1 + \gamma_5)^2 ({\displaystyle{\not}p}_{h} - {\displaystyle{\not}p}_1+M_{N_j}) \right]
\label{bart1}
\\
&=&
8 {\Big \{} (p_1 \cdot p_h) \left[ 4 (p_2 \cdot p_{\pi})^2 + 2 m_{\pi}^2 (p_2 \cdot p_{\pi}) - 2 m_{\pi}^2 m_{\ell_2}^2 \right]
\nonumber\\ && +
\left[ - 2  (p_1 \cdot p_{\pi})(p_2 \cdot p_{\pi}) + m_{\pi}^2 (p_1 \cdot p_2) \right] \left[ (p_h - p_1)^2 - M_{N_i} M_{N_j} \right] {\Big \}}.
\label{bart2} \eea \ees
When we use the on-shell conditions for the initial and final state particles, we obtain
\bes
\label{Texpr}
\bea
\lefteqn{
{\overline T}(p_N^2)_{ij} \equiv  \frac{1}{8} {\bar t}_{i j} =
{\Big \{} \frac{1}{2} (M_h^2 - p_N^2) \left[ (p_N^2 - m_{\ell_2}^2 - m_{\pi}^2) (p_N^2  - m_{\ell_2}^2) - 2 m_{\pi}^2 m_{\ell_2}^2 \right]
}
\label{tbulk} \\
&& +  \frac{1}{2} \left[ p_N^2 - M_{N_i} M_{N_j} \right]  \left[ 2 (p_1 \cdot p_2) (p_N^2 - m_{\ell_2}^2) - (M_h^2 - p_N^2) (p_N^2 - m_{\ell_2}^2 - m_{\pi}^2) \right]
{\Big \}}.
\label{tcorr}
\eea \ees
When we take into account the near degeneracy of the two intermediate neutrinos
\bes
\label{massdeg}
\bea
M_{N_1} &\equiv & M_N; \quad \Delta M_N \equiv M_{N_2} - M_{N_1};
\label{MNdef} \\
&& 0 < \Delta M_N (\lesssim \Gamma_{N_j}) \ll M_N,
\label{deg} \eea \ees
the quadratic forms of the intermediate neutrino propagators, in the considered limiting case (\ref{deg}), can be written as
\bes
\label{PP}
\bea
P_j P_j^{\ast} &=& \frac{\pi}{M_N \Gamma_{N_j}} \delta(p_N^2 - M_N^2) \quad (j=1,2),
\label{PjPj} \\
{\rm Im} (P_1 P_2^{\ast}) & = & \frac{\eta(y)}{y} \frac{\pi}{M_N \Gamma_{N}} \delta(p_N^2 - M_N^2),
\label{ImP1P2} \\
{\rm Re} (P_1 P_2^{\ast}) & = & \delta(y) \frac{\pi}{M_N \Gamma_{N}} \delta(p_N^2 - M_N^2),
\label{ReP1P2}
\eea \ees
where
\bes
\label{yetdel}
\bea
y & = &\frac{\Delta M_N}{\Gamma_N},  \qquad \Gamma_N = \frac{1}{2} ( \Gamma_{N_1} + \Gamma_{N_2} ),
\label{yGN} \\
\frac{\eta(y)}{y} & = & \frac{y}{1+y^2},
\label{eta} \\
\delta(y) &=& \frac{1}{1 + y^2}.
\label{del} \eea \ees
The expressions (\ref{eta})-(\ref{del}) were obtained numerically in rare decay processes of pseudoscalar mesons via almost degenerate neutrinos $N_j$ ($j=1,2$) in Refs.~\cite{GCCSKJZS1,GCCSKJZS2}, and a derivation of formula (\ref{eta}) was presented in Ref.~\cite{symm} (App. 6 there). Here we present in Appendix \ref{app:delta} a derivation of formula (\ref{del}).

In order to obtain the decay widths of the processes for the amplitudes ${\cal T}_{\mp}$, we integrate their squares over the final phase space
\bea
d \Gamma(h \to \nu_k \ell_2^{\mp} \pi^{\pm}) & = & \frac{1}{ 2 M_h (2 \pi)^5} |{\cal T}_{\mp}|^2 \; d_3,
\label{dGamma1} \eea
where $d_3$ denotes the differential of the integration over the phase space of the three final particles
\be
\label{d3}
d_3  =   d_2(h \to \nu_k N_j) \; d p_N^2 \; d_2(N_j \to \ell_2 \pi).
\ee
Here, the three-particle differentials $d_2$ are
\bes
\label{d2s}
\bea
\frac{d_2(h \to \nu_k N_j)}{d \Omega_{{\hat p}^{'}_N}} &=& \frac{1}{8} \lambda^{1/2} \left(1, 0, \frac{p_N^2}{M_h^2} \right),
\label{d2h} \\
\frac{d_2(N \to \ell_2 \pi)}{d \Omega_{{\hat p}^{''}_2}} &=& \frac{1}{8} \lambda^{1/2} \left(1, \frac{m_{\ell_2}^2}{p_N^2}, \frac{m_{\pi}^2}{p_N^2} \right),
\label{d2N} \eea \ees
where ${\hat p_N}'$ is the direction of $N_j$ in the Higgs-rest system ($\Sigma'$) , and ${\hat p^{''}_2}$ is the direction of $\ell_2^{\mp}$ in the $N_j$-rest system ($\Sigma''$),\footnote{
We will denote the lab system as the unprimed system ($\Sigma$).}
and $\lambda^{1/2}(x,y,z)$ is square root of the function
\be
\lambda(x,y,z) = x^2 + y^2 + z^2 - 2 x y - 2 y z - 2 z x.
\label{lamb} \ee
We also take into account that the ratios appearing in Eqs.~(\ref{Tsq}) before the integral there are practically equal to unity
\be
\frac{\left( 1 - \frac{m_{\pi}^2}{M_W^2} \right)^2 }{ \left[ \left( 1 -  \frac{m_{\pi}^2}{M_W^2} \right)^2 + \left( \frac{\Gamma_W}{M_W} \right)^2 \right]} = 1 + {\cal O}(10^{-3}) \approx 1.
\label{ratapp} \ee
Furthermore, when taking into account the expressions (\ref{PP}) in (\ref{Tsq}), and denoting
\be
U_{\ell_2 N_j} = |U_{\ell_2 N_j}| e^{i \phi_j} \quad (j=1,2); \quad \Delta \phi \equiv \phi_2 - \phi_1,
\label{delphi} \ee
the squared decay amplitudes obtain the following form:
\bea
|{\cal T}_{\mp}|^2 & = & 8 K^2 T  \delta(p_N^2-M_N^2) \frac{\pi}{M_N}
{ \bigg \{} \frac{1}{\Gamma_{N_1}} ({\cal A}^{(k)}_1)^2  |U_{\ell_2 N_1}|^2 +
\frac{1}{\Gamma_{N_2}}  ({\cal A}^{(k)}_2)^2 |U_{\ell_2 N_2}|^2
\nonumber\\ &&
+
\frac{2}{\Gamma_N} {\cal A}^{(k)}_1 {\cal A}^{(k)}_2 |U_{\ell_2 N_1}| |U_{\ell_2 N_2}|
\left[ \delta(y) \cos(\Delta \phi) \pm \frac{ \eta(y)}{y} \sin ( \Delta \phi) \right] {\bigg \}}.
\label{Tsq2}
\eea
The Dirac-delta $\delta(p_N^2-M_N^2)$ above implied that the expression ${\overline T}(p_N^2)_{i j}$ of Eq.~(\ref{Texpr}) got simplified (${\overline T}(p_N^2)_{i j} \mapsto T$), i.e., $p_N^2 \mapsto M_N^2$ and the second part of this expression disappears
\be
{\overline T}(p_N^2)_{i j} \mapsto T = \frac{1}{2} (M_h^2 - M_N^2) \left[ (M_N^2 - m_{\ell_2}^2 - m_{\pi}^2)  (M_N^2  - m_{\ell_2}^2) - 2 m_{\pi}^2 m_{\ell_2}^2 \right].
\label{texpr} \ee
The integration $d p_N^2$ in the differential $d_3$ Eq.~(\ref{d3}) can be immediately performed in Eq.~(\ref{dGamma1}), because of the Dirac-delta factor $\delta(p_N^2-M_N^2)$ in the integrand, leading to
\bea
\frac{d \Gamma(h \to \nu_k \ell_2^{\mp} \pi^{\pm})}{d \Omega_{{\hat p}_N} d \Omega_{{\hat p}^{'}_2}} & = &
\frac{1}{ 2 M_h (2 \pi)^5}  \frac{1}{8^2} \lambda^{1/2} \left(1, 0, \frac{M_N^2}{M_h^2} \right)   \lambda^{1/2} \left(1, \frac{m_{\ell_2}^2}{M_N^2}, \frac{m_{\pi}^2}{M_N^2} \right)
\nonumber\\ && \times
 8 K^2 T \frac{\pi}{M_N \Gamma_N}
{\bigg \{} \frac{\Gamma_N}{\Gamma_{N_1}} ({\cal A}^{(k)}_1)^2  |U_{\ell_2 N_1}|^2 +
\frac{\Gamma_N}{\Gamma_{N_2}}  ({\cal A}^{(k)}_2)^2 |U_{\ell_2 N_2}|^2
\nonumber\\ &&
+
2 {\cal A}^{(k)}_1 {\cal A}^{(k)}_2 |U_{\ell_2 N_1}| |U_{\ell_2 N_2}|
\left[ \delta(y) \cos(\Delta \phi) \pm \frac{ \eta(y)}{y} \sin ( \Delta \phi) \right] {\bigg \}}
\label{dGdOm1}
\eea
Integration over $ d \Omega_{{\hat p}^{''}_2}$ and $d \Omega_{{\hat p}^{'}_N}$ gives factor $(4 \pi)^2$ because the integrand has no dependence on these directions.
This then gives us the decay width of the considered rare Higgs decay
\bea
\Gamma(h \to \nu_k \ell_2^{\mp} \pi^{\pm}) & = &
{\overline \Gamma}(h \to \nu_k \ell_2 \pi)
{\bigg \{} \frac{\Gamma_N}{\Gamma_{N_1}} ({\cal A}^{(k)}_1)^2  |U_{\ell_2 N_1}|^2 +
\frac{\Gamma_N}{\Gamma_{N_2}}  ({\cal A}^{(k)}_2)^2 |U_{\ell_2 N_2}|^2
\nonumber\\ &&
+
2 {\cal A}^{(k)}_1 {\cal A}^{(k)}_2 |U_{\ell_2 N_1}| |U_{\ell_2 N_2}|
\left[ \delta(y) \cos(\Delta \phi) \pm \frac{ \eta(y)}{y} \sin ( \Delta \phi) \right] {\bigg \}},
\label{Gd1}
\eea
where ${\overline \Gamma}(h \to \nu_k \ell_2 \pi)$ is the canonical decay width expression
\bea
{\overline \Gamma}(h \to \nu_k \ell_2 \pi) &=& \frac{K^2}{64 \pi^2} \frac{M_h M_N^3}{\Gamma_N} \left( 1 - \frac{M_N^2}{M_h^2} \right)^2 \left[ 1 - x_{\pi} - 2 x_{\ell_2} - x_{\ell_2} (x_{\pi} - x_{\ell_2}) \right] \lambda^{1/2}(1,x_{\ell_2},x_{\pi}),
\label{barG} \eea
where we used the notations
\be
x_{\ell_2} = \frac{m_{\ell_2}^2}{M_N^2}, \qquad x_{\pi} = \frac{m_{\pi}^2}{M_N^2}.
\label{xs} \ee
It can be checked that the canonical expression (\ref{barG}) can be written in the factorised form
\be
 {\overline \Gamma}(h \to \nu_k \ell_2 \pi) =
 {\overline \Gamma}(h \to \nu_k N) \frac{{\overline \Gamma}(N \to \ell_2 \pi)}{\Gamma_N},
\label{fact} \ee
where the two factors
\bes
\label{barGs}
\bea
{\overline \Gamma}(h \to \nu_k N) & = & \frac{M_h}{4 \pi} \left( 1 - \frac{M_N^2}{M_h^2} \right)^2
\label{barGh} \\
{\overline \Gamma}(N \to \ell_2 \pi) & = & \frac{K^2}{16 \pi} M_N^3 \left[ 1 - x_{\pi} - 2 x_{\ell_2} - x_{\ell_2} (x_{\pi} - x_{\ell_2}) \right] \lambda^{1/2}(1,x_{\ell_2},x_{\pi})
\label{barGN} \eea \ees
are the canonical decay widths for the decay processes $h \to \nu_k N$ and $N \to \ell_2 \pi$, respectively, i.e., when the corresponding couplings are $- \sqrt{2} h {\bar \nu}_k N$ and $(-g \sqrt{2}/4) ( {\bar \ell}_2 {\displaystyle{\not}W} N_L + {\rm h.c.})$, i.e., the cases of one neutrino of mass $M_N$ and the coupling parameters ${\cal A}^{(k)}_N \mapsto 1$ and $(-g \sqrt{2}/4) U_{\ell_2 N} \mapsto (-g \sqrt{2}/4)$.

The formula (\ref{Gd1}) [with notations Eqs.~(\ref{barG})-(\ref{xs})] was obtained here in the near degeneracy case Eqs.~(\ref{massdeg}), by using the formulas (\ref{PP})-(\ref{yetdel}) which hold in this case. In Appendix \ref{app:Gd1exact} we present, for comparison, the more general formulas for the quantity $\Gamma(h \to \nu_k \ell_2^{\mp} \pi^{\pm})$, which are not restricted to the near degeneracy case.

\begin{figure}[htb] 
\centering\includegraphics[width=100mm]{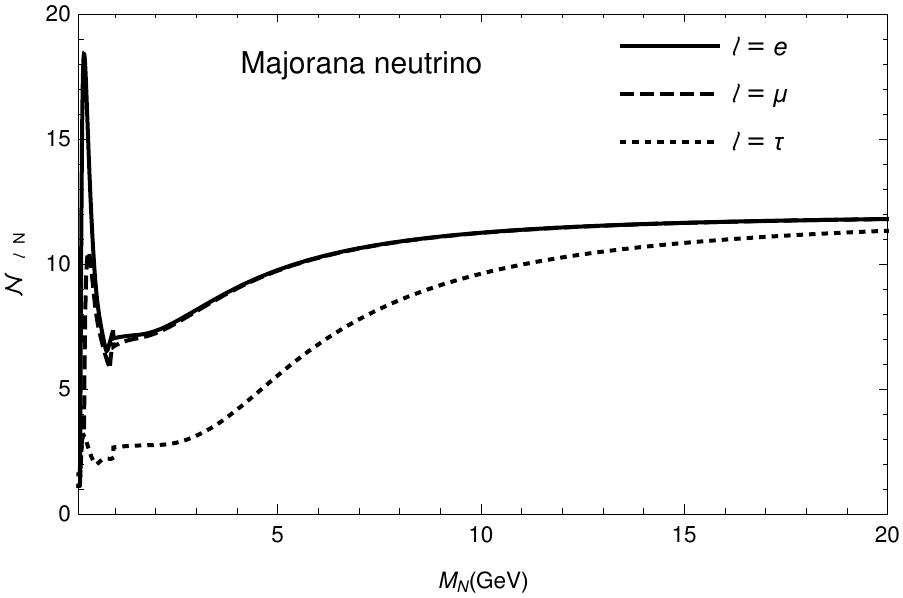}
\caption{\footnotesize The factors ${\cal N}_{\ell N}$ appearing in the Majorana neutrino decay width Eqs.~(\ref{GammaNj})-(\ref{calKj}), as a function of the mass $M_N$ of the Majorana neutrino $N_j$.}
\label{FigGN}
\end{figure}

The canonical decay widths (${\overline \Gamma}$) refer to the decay cases where the heavy-light neutrino mixing coefficients $|U_{\ell N_j}|^2$ [cf.~Eq.~(\ref{mix})] are unity. On the other hand, the total decay width $\Gamma_{N_j}$ of the neutrino $N_j$ [for $\Gamma_N$ cf.~Eq.~(\ref{yGN})] is the true decay width, which contains these heavy-light mixing coefficients (cf.~\cite{GCCSKJZS2,symm} and references therein for details)
\be
\Gamma_{N_j} = {\widetilde {\cal K}}_j(M_N) {\overline \Gamma}_N(M_N),
\label{GammaNj} \ee
where
\be
 {\overline \Gamma}_N(M_N) \equiv \frac{G_F^2 M_N^5}{96 \pi^3},
\label{bGN} \ee
and the factor ${\widetilde {\cal K}}_j$ ($j=1,2$) contains the heavy-light mixing coefficients
\be
 {\widetilde {\cal K}}_j(M_N) = {\cal N}_{e N} |U_{e N_j}|^2 +{\cal N}_{\mu N} |U_{\mu N_j}|^2 +{\cal N}_{\tau N} |U_{\tau N_j}|^2 \qquad (j=1,2).
\label{calKj} \ee
The factors ${\cal N}_{\ell N} = {\cal N}_{\ell N}(M_N)$  ($\ell=e, \mu, \tau$) are the effective mixing coefficients, they are dimensionless numbers $\sim 10^1$, and were evaluated in \cite{GCCSKJZS2}.  In Fig.~\ref{FigGN} we present the factor ${\cal N}_{\ell N}(M_N)$ as a function of the mass $M_N$ of the Majorana neutrinos $N_j$.
Incidentally, we can see in Fig.~\ref{FigGN} that in the low mass regime ($M_N \approx 0.2$-$0.4$ GeV) the factors ${\cal N}_{\ell N}$ for light leptons $\ell=e, \mu$ have a strong variation and a local maximum. This occurs because for such masses $M_N$ the N decay channels which become important are those to the single light pseudoscalar and vector mesons, for which $\Gamma$ is proportional to $M_N^3$; for the other decay channels (to pure lepton channels and to open quark channels) we have $\Gamma$ proportional to $M_N^5$. Since in the total decay width we factor out $M_N^5$ (and not $M_N^3$), cf.~Eqs.~(\ref{GammaNj})-(\ref{bGN}), the factor ${\cal N}_{\ell N}(M_N)$ becomes roughly proportional to $1/M_N^2$ in the interval $0.3 \ {\rm GeV} < M_N < 1 \ {\rm GeV}$, i.e., despite the appearance of new channels it falls there when $M_N$ increases.

\begin{figure}[htb] 
\centering\includegraphics[width=100mm]{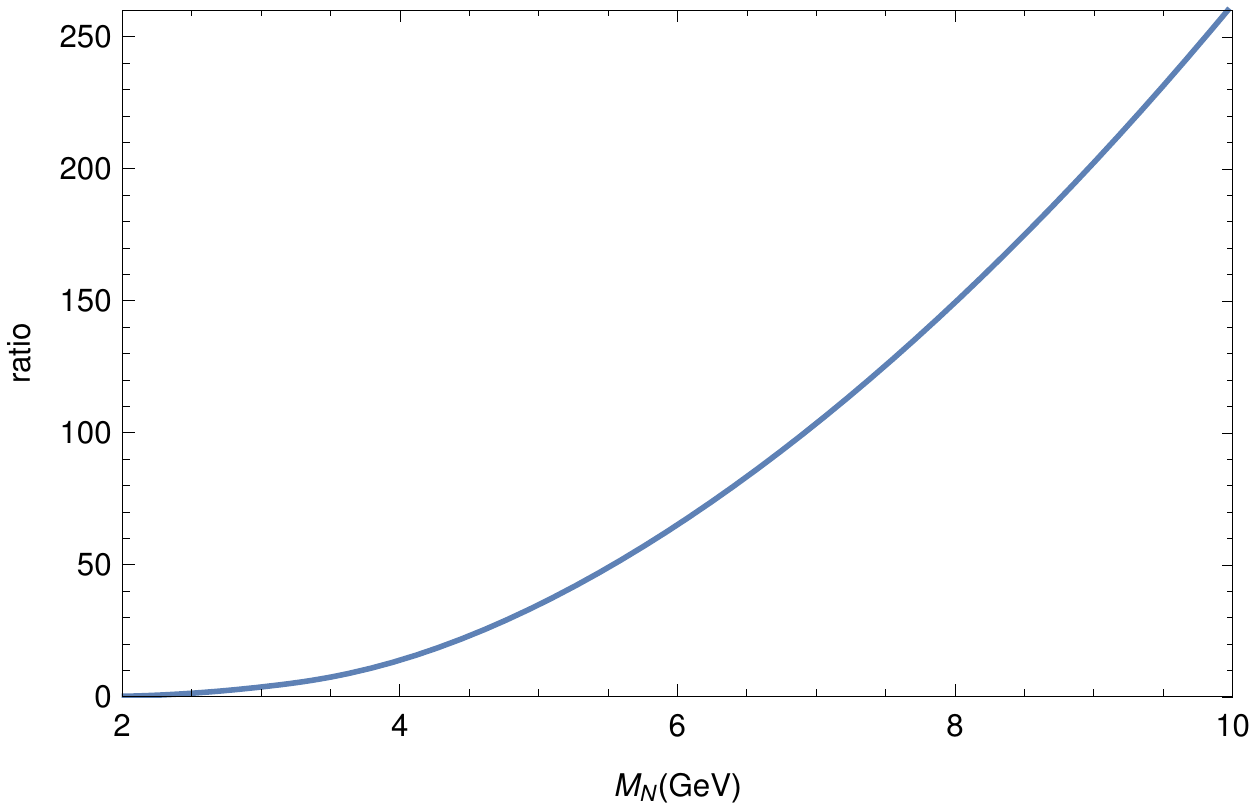}
\caption{\footnotesize The enhancement ratio ${\overline \Gamma}(N \to \ell_2^{-} U {\bar D})/{\overline \Gamma}(N \to \ell_2^{-} \pi^+)$, as a function of mass $M_N$ of Majorana neutrino $N_j$, for $\ell_2=\tau$. The sum over $U=u,c$ and $D=d, s$ is implied whenever kinematically allowed. The used values of the quark masses are: $m_u=m_d=3.5$ MeV; $m_s=105$ MeV; $m_c=1.27$ GeV.}
\label{Figrat}
\end{figure}

It turns out that the International Linear Collider (ILC) or a similar future Higgs factory will have almost no hadronic uncertainty. Therefore, at the second vertex of the considered process Fig.~\ref{Fighdec} we can consider, instead of the specific decay $N \to \ell_2^{\mp} \pi^{\pm}$, the inclusive (i.e., open) quark channel decays $N \to \ell_2^{-} U {\bar D}$   and $N \to \ell_2^{+} {\bar U} D$, where $U = u, c$ and $D=d, s$.\footnote{$U {\bar D}$ and ${\bar U} D$ appear in practice as two jets ($jj$), which in the considered cases have nonzero total charge. Even though the jets may not be identified, the detection of the emission of the charged lepton $\ell_2^{\mp}$ at the second vertex ensures that the produced jets $jj$ have nonzero ($\pm$) charge.} This then means that the result (\ref{Gd1}) can be immediately extended to the form\footnote{The simplicity of this extension is related to the effective on-shellness of the (quasidegenerate) intermediate neutrinos $N_j$ ($j=1,2$) as reflected by the Dirac delta function factor $\delta(P_N^2 - M_N^2)$ in Eqs.~(\ref{PP}).}
\bea
\Gamma(h \to \nu_k \ell_2^{\mp} U D) & = &
{\overline \Gamma}(h \to \nu_k \ell_2 U D)
{\bigg \{} \frac{\Gamma_N}{\Gamma_{N_1}} ({\cal A}^{(k)}_1)^2  |U_{\ell_2 N_1}|^2 +
\frac{\Gamma_N}{\Gamma_{N_2}}  ({\cal A}^{(k)}_2)^2 |U_{\ell_2 N_2}|^2
\nonumber\\ &&
+
2 {\cal A}^{(k)}_1 {\cal A}^{(k)}_2 |U_{\ell_2 N_1}| |U_{\ell_2 N_2}|
\left[ \delta(y) \cos(\Delta \phi) \pm \frac{ \eta(y)}{y} \sin ( \Delta \phi) \right] {\bigg \}},
\label{Gd1ext}
\eea
where\footnote{The left-hand side of Eq.~(\ref{Gd1ext}) has simplified notation; it means that for $\ell_2^{-}$ we have $U {\bar D}$, and for $\ell_2^{+}$ we have ${\bar U} D$.} the sum over $U=u,c$ and $D=d,s$ is implied whenever kinematically allowed, and the canonical decay width ${\overline \Gamma}(h \to \nu_k \ell_2 U D)$ has the factorised form analogous to Eq.~(\ref{fact}), but with the second canonical factor ${\overline \Gamma}(N \to \ell_2 \pi)$ replaced by the canonical factor ${\overline \Gamma}(N \to \ell_2 U D)$
\bes
\label{factbarGNext}
\bea
{\overline \Gamma}(h \to \nu_k \ell_2 U D) &=&
 {\overline \Gamma}(h \to \nu_k N) \frac{{\overline \Gamma}(N \to \ell_2 U D)}{\Gamma_N},
\label{factext} \\
{\overline \Gamma}(N \to \ell_2 U D) &=& \frac{G_F^2}{64 \pi^3} M_N^5 |V_{UD}|^2 J_1(x_{\ell_2}, x_U, x_D),
\label{barGNext} \eea \ees
where $x_{\ell_2}$ was defined in Eq.~(\ref{xs}), $x_U = m_U^2/M_N^2$, $x_D=m_D^2/M_N^2$, the factor $V_{UD}$ is the corresponding CKM matrix element, and the (one-loop) kinematical expression $J_1$ is \cite{GKS,HKS} (cf.~also \cite{GCCSKJZS1,symm})
\be
J_1(x_{\ell_2},x_U,x_D) = 12 \int_{s_{\rm min}}^{s_{\rm max}} \frac{ds}{s} (s -x_{\ell_2} - x_U) (1 + x_D - s) \lambda^{1/2}(s,x_{\ell_2},x_U) \lambda^{1/2}(1,s,x_D),
\label{J1} \ee
where $s_{\rm min} = (\sqrt{x_{\ell_2}}+ \sqrt{x_U})^2$ and $s_{\rm max}=(1 - \sqrt{x_D})^2$.
This effect, for the masses $M_N \approx 5$-$10$ GeV, increases the decay width $\Gamma(h \to \nu_k \ell_2 U D)$ Eq.~(\ref{Gd1ext}), with respect to  $\Gamma(h \to \nu_k \ell_2 \pi)$ Eq.~(\ref{Gd1}), by a factor $\sim 10^2$. In Fig.~\ref{Figrat} we present this enhancement as a function of the mass $M_N$.

Until now we assumed that the intermediate neutrinos $N_j$ both decay within the detector, i.e., that the detector is infinitely large. However, since the detector has a finite length $L_{\rm det}$ ($\sim 1$ m), we have to exclude from the decay width the decays where the distance $L$ between the two vertices of the decay process is larger than the detector width, $L > L_{\rm det}$. This leads us to an effective decay width $\Gamma_{\rm eff}(L)$ as a function of the maximal length $L$ between the two vertices. The corresponding differential $d \Gamma_{\rm eff}(L) =  \Gamma_{\rm eff}(L+dL)-\Gamma_{\rm eff}(L)$ is then
\bea
d \Gamma_{\rm eff}(h \to \nu_k \ell_2^{\mp} U D; L) & = &
{\overline \Gamma}(h \to \nu_k \ell_2 U D)  {\bigg \{} \frac{\Gamma_N}{\Gamma_{N_1}} d P_{N_1}(L) ({\cal A}^{(k)}_1)^2  |U_{\ell_2 N_1}|^2  + \frac{\Gamma_N}{\Gamma_{N_2}} d P_{N_2}(L) ({\cal A}^{(k)}_2)^2  |U_{\ell_2 N_2}|^2  +
\nonumber\\ &&
+
d P_{N}(L) 2 {\cal A}^{(k)}_1 {\cal A}^{(k)}_2 |U_{\ell_2 N_1}| |U_{\ell_2 N_2}|
\left[ \delta(y) \cos(\Delta \phi) \pm \frac{ \eta(y)}{y} \sin ( \Delta \phi) \right] {\bigg \}}
\label{dGeff1} \eea
where $dP_{N_j}(L) = P_{N_j}(L+d L) - P_{N_j}(L)$, and $P_{N_j}(L)$ is the probability that the propagating $N_j$ neutrino decays at a distance $L$ from its birth vertex
\bes
\label{PNjs}
\bea
P_{N_j}(L) &=& 1 - \exp \left( - \frac{L \Gamma_{N_j}}{\beta_N \gamma_N} \right),
\label{PNj} \\
d P_{N_j}(L) &=& \frac{\Gamma_{N_j}}{\beta_N \gamma_N} \exp \left( - \frac{L \Gamma_{N_j}}{\beta_N \gamma_N} \right) d L,
\label{dPNj} \eea \ees
where $\Gamma_{N_j}$ is the (earlier encountered) total decay width of $N_j$, and $\gamma_N = (1 - \beta_N)^{-1/2}$ is the Lorentz lab time dilation factor ($\beta_N$ is the speed of $N_j$ in the Lorentz frame). It is not clear which value of $\Gamma_{N_j}$ to use in $d P_{N}(L)$ at the $N_1$-$N_2$ overlap contributions $\propto \delta(y), \eta(y)/y$ in Eq.~(\ref{dGeff1}). We will use at this point the average value $\Gamma_N$ (\ref{yGN}); however, this will not matter, as we will later assume that $\Gamma_{N_1} = \Gamma_{N_2}$ (i.e., that $|U_{\ell_s N_1}|= |U_{\ell_s N_2}|$ for all $\ell_s$).

\section{Inclusion of oscillation effects}
\label{sec:Gosc}

As argued in Ref.~\cite{oscGCetal}, based on the approach of Ref.~\cite{CGL}, the effects of oscillation of the (quasidegenerate) $N_1$ and $N_2$ neutrinos during their propagation between the two vertices of the decay process lead to the following replacements:
\bes
\label{oscrepl}
\bea
{\cal T}_{-}: \; U_{\ell_2 N_j}       & \mapsto & U_{\ell_2 N_j}        \exp(- i p_{N_j} \cdot z)
\label{Tmoscrepl} \\
{\cal T}_{+}: \; U_{\ell_2 N_j}^{\ast} & \mapsto & U_{\ell_2 N_j}^{\ast} \exp(- i p_{N_j} \cdot z) \qquad (j=1,2).
\label{Tposcrepl}
\eea \ees
Here, $z$ is the 4-vector of distance between the two vertices, $z=(t,0,0,L)$, and $p_{N_1}$ and $p_{N_2}$ are the momenta of the two types of neutrinos. These two momenta slighly differ from each other due to the (small) mass difference between the two neutrinos. The relation is independent of the charge channel of the particles produced at the second vertex ($\ell_2^- U {\bar D}$ or $\ell_2^+ {\bar U} D$). As argued in \cite{CGL,oscGCetal}, the resulting difference in phases can be expressed as
\be
(p_{N_2}-p_{N_1}) \cdot z = 2 \pi \frac{L}{L_{\rm osc}},
\label{difosc} \ee
where the effective oscillation length is
\be
L_{\rm osc} = \frac{2 \pi \beta_N \gamma_N}{\Delta M_N}, \qquad
\frac{2 \pi}{L_{\rm osc}} = y \frac{\Gamma_N}{\beta_N \gamma_N} .
\label{Losc} \ee
Here, $\beta_N$ is the speed of the on-shell neutrinos $N_j$ in the lab (practically equal for $N_1$ and $N_2$) and $\gamma_N = 1/\sqrt{1 - \beta_N^2}$ is the corresponding Lorentz factor.

The two amplitudes ${\cal T}_{\mp}$ with oscillation effects can be schematically described as
\bes
\label{osceff1}
\bea
{\cal T}(h \to \nu_k \ell_2^- U {\bar D}) &\sim& {\cal A}^{(k)}_1 U_{\ell_2 N_1} \exp(- i p_{N1} \cdot z) + {\cal A}^{(k)}_2 U_{\ell_2 N_2} \exp(- i p_{N2} \cdot z),
\label{Tmosc} \\
{\cal T}(h \to \nu_k \ell_2^+ {bar U} D) &\sim& {\cal A}^{(k)}_1 U_{\ell_2 N_1}^{\ast} \exp(- i p_{N1} \cdot z) + {\cal A}^{(k)}_2 U_{\ell_2 N_2}^{\ast} \exp(- i p_{N2} \cdot z).
\label{Tposc} \eea \ees
Using the approach described in \cite{oscGCetal},\footnote{In Ref.~\cite{oscGCetal}, the LNV decays $B^{\pm} \to \mu^{\pm} N_j \to \mu^{\pm} e^{\pm} \pi^{\mp}$ were considered, and the $N_1$-$N_2$ overlap effects were considered in a simplified schematical way, i.e., the product of propagators $P_1 P_2^{\ast}$ was simply taken equal to $|P_1|^2$, this then corresponding to $\delta(y) \mapsto 1$ and $\eta(y)/y \mapsto 0$, cf.~Eqs.~(\ref{PP}).}
the interference effects in the square of the schematic amplitude (\ref{osceff1}) then result in an oscillating term $\propto \cos( 2 \pi (L/L_{\rm osc}) \mp \Delta \phi)$
\bea
d \Gamma_{\rm eff}(h \to \nu_k \ell_2^{\mp} U D; L) &=&
{\overline \Gamma}(h \to \nu_k \ell_2 U D) {\bigg \{}
\frac{\Gamma_N}{\Gamma_{N_1}} d P_{N_1}(L) ({\cal A}^{(k)}_1)^2  |U_{\ell_2 N_1}|^2  + \frac{\Gamma_N}{\Gamma_{N_2}} d P_{N_2}(L) ({\cal A}^{(k)}_2)^2  |U_{\ell_2 N_2}|^2
\nonumber\\ &&
+ d P_{N}(L) 2 {\cal A}^{(k)}_1 {\cal A}^{(k)}_2 |U_{\ell_2 N_1}| |U_{\ell_2 N_2}|
\cos\left( 2 \pi \frac{L}{L_{\rm osc}} \mp \Delta \phi \right)  {\bigg \}}.
\label{dGeff2} \eea
Since the amplitudes were taken in a schematic form, cf.~Eq.~(\ref{osceff1}), the factor in front of the cosinus in Eq.~(\ref{dGeff2}) has a naive (schematic) $N_1$-$N_2$ overlap because in the products of propagators $P_1 P_2^{\ast} \mapsto |P_1|^2$ ($\approx |P_2|^2$) was used [cf.~Eqs.~(\ref{PP})].
The oscillation term of the type of Eq.~(\ref{dGeff2}) can be naively combined with the $N_1$-$N_2$ overlap terms of Eq.~(\ref{dGeff1}), by adding the two terms
\bea
\lefteqn{
d \Gamma_{\rm eff}(h \to \nu_k \ell_2^{\mp} U D; L) =
{\overline \Gamma}(h \to \nu_k \ell_2 U D) {\bigg \{}
\frac{\Gamma_N}{\Gamma_{N_1}} d P_{N_1}(L) ({\cal A}^{(k)}_1)^2  |U_{\ell_2 N_1}|^2  + \frac{\Gamma_N}{\Gamma_{N_2}}  d P_{N_2}(L) ({\cal A}^{(k)}_2)^2  |U_{\ell_2 N_2}|^2  +
}
\nonumber\\ &&
+
d P_{N}(L) 2 {\cal A}^{(k)}_1 {\cal A}^{(k)}_2 |U_{\ell_2 N_1}| |U_{\ell_2 N_2}|
\left[ \delta(y) \cos(\Delta \phi) \pm \frac{ \eta(y)}{y} \sin ( \Delta \phi)  + \cos\left( 2 \pi \frac{L}{L_{\rm osc}} \mp \Delta \phi \right) \right] {\bigg \}}
,
\label{dGeff3} \eea
This approximate approach, which thus contains the sum of the overlap effects without the oscillation and the oscillation effects ``without'' the overlap (i.e., with naive overlap), was used in Ref.~\cite{CPVBelle} in the LNV decays $B^{\pm} \to D^0 \ell_1^{\pm} N_j \to  D^0 \ell_1^{\pm} \ell_2^{\pm} \pi^{\pm}$.

Nonetheless, a more systematic approach of inclusion of the mentioned oscillation effects can be performed, by making replacements (\ref{oscrepl}) in the full amplitudes ${\cal T}_{\mp}$ in Eqs.~(\ref{Tm}) and (\ref{Tp}).\footnote{It is understood that we subsequently replace $\pi^+$ by $U {\bar D}$ and $\pi^-$ by ${\bar U} D$; this then redefines the expression $T$ of Eq.~(\ref{texpr}) accordingly [cf.~Eqs.~(\ref{Gd1ext})-(\ref{factbarGNext})].} This then leads to the absolute squares of these amplitudes in the following form:
\bea
|{\cal T}_{-}|^2 & = & 8 K^2 T \delta(p_N^2-M_N^2) \frac{\pi}{M_N \Gamma_N}
{ \bigg \{} \frac{\Gamma_{N}}{\Gamma_{N_1}} ({\cal A}^{(k)}_1)^2  |U_{\ell_2 N_1}|^2 +
\frac{\Gamma_{N}}{\Gamma_{N_2}}  ({\cal A}^{(k)}_2)^2 |U_{\ell_2 N_2}|^2
\nonumber\\ &&
+
2  {\cal A}^{(k)}_1 {\cal A}^{(k)}_2
{\Big [} U_{\ell_2 N_1} U_{\ell_2 N_2}^{\ast} \exp \left( +i (p_{N_2} - p_{N_1}) \cdot z \right) \left( \delta(y) +i \frac{\eta(y)}{y} \right)
\nonumber\\ &&
  + U_{\ell_2 N_2} U_{\ell_2 N_1}^{\ast} \exp \left( - i (p_{N_2} - p_{N_1}) \cdot z \right) \left( \delta(y) -i \frac{\eta(y)}{y} \right) {\Big ]} {\bigg \}}.
\label{Tmsqrosc1} \eea
Similar expression can be obtained for $|{\cal T}_{+}|^2$.
After straightforward algebra and using our notations, we obtain:
\bea
|{\cal T}_{\mp}|^2 & = & 8 K^2 T \delta(p_N^2-M_N^2) \frac{\pi}{M_N \Gamma_N}
{ \bigg \{} \frac{\Gamma_{N}}{\Gamma_{N_1}} ({\cal A}^{(k)}_1)^2  |U_{\ell_2 N_1}|^2 +
\frac{\Gamma_{N}}{\Gamma_{N_2}}  ({\cal A}^{(k)}_2)^2 |U_{\ell_2 N_2}|^2
\nonumber\\ &&
+ 2  {\cal A}^{(k)}_1 {\cal A}^{(k)}_2 |U_{\ell_2 N_1}| |U_{\ell_2 N_2}|
\left[ \delta(y) \cos \left( 2 \pi \frac{L}{L_{\rm osc}} \mp \Delta \phi \right) +
  \frac{\eta(y)}{y} \sin\left( 2 \pi \frac{L}{L_{\rm osc}} \pm \Delta \phi \right)
  \right] {\bigg \}}.
\label{Tmsqrosc2} \eea
This then leads to the final form of the effective $d \Gamma_{\rm eff}$
\bea
\lefteqn{
d \Gamma_{\rm eff}(h \to \nu_k \ell_2^{\mp} U D; L) =
{\overline \Gamma}(h \to \nu_k \ell_2 U D) {\bigg \{}
\frac{\Gamma_N}{\Gamma_{N_1}} d P_{N_1}(L) ({\cal A}^{(k)}_1)^2  |U_{\ell_2 N_1}|^2  +
\frac{\Gamma_N}{\Gamma_{N_2}} d P_{N_2}(L) ({\cal A}^{(k)}_2)^2  |U_{\ell_2 N_2}|^2 +
}
\nonumber\\ &&
+
d P_{N}(L) 2 {\cal A}^{(k)}_1 {\cal A}^{(k)}_2 |U_{\ell_2 N_1}| |U_{\ell_2 N_2}|
\left[ \delta(y) \cos\left( 2 \pi \frac{L}{L_{\rm osc}} \mp \Delta \phi \right)
  + \frac{ \eta(y)}{y} \sin\left( 2 \pi \frac{L}{L_{\rm osc}} \pm \Delta \phi \right) \right]  {\bigg \}}.
\label{dGeff4} \eea
We point out that in this result, the overlap and oscillation effects appear strongly intertwined. The usual overlap expression (\ref{dGeff1}) is obtained from Eq.~(\ref{dGeff4}) in the limit of $L/L_{\rm osc} \to 0$ (no oscillation). On the other hand, the oscillation result Eq.~(\ref{dGeff2}) where the schematic (naive) overlap was used is obtained from Eq.~(\ref{dGeff4}) in the limit $y \to 0$ [then: $\delta(y) \to 1$ and $\eta(y)/y \to 0$]. We can see from the expression (\ref{dGeff4}) that the combined formula (\ref{dGeff3}) is naive, in the sense that it contains the overlap and the oscillation effects as a sum of the two corresponding terms, i.e., not intertwined.

From now on, we will adopt a simplifying assumption about the heavy-light mixing, namely that
\be
|U_{\ell N_1}| = |U_{\ell N_2}|  \; (= |U_{\ell N}|) \qquad (\ell = e, \mu, \tau).
\label{simpl} \ee
This then implies that the total decay widths of $N_1$ and $N_2$ are equal\footnote{Cf.~formulas given in Refs.~\cite{GCCSKJZS1,GCCSKJZS2,symm} for the heavy neutrino decay widths.}, and thus the probability differentials $d P_{N_j}(L)$ in Eq.~(\ref{dGeff4}) are equal [cf.~Eq.~(\ref{dPNj})]
\be
\Gamma_{N_1} = \Gamma_{N_2} = \Gamma_{N}; \qquad dP_{N_j}(L) =dP_N(L) \quad (j=1,2).
\label{dPN} \ee
Under these assumptions, the formula (\ref{dGeff4}) simplifies somewhat with regard to $L$-dependence, and integration over $dL'$ from $L'=0$ to $L'=L$ gives
[using also Eq.~(\ref{dPNj}) and Eqs.~(\ref{eta})-(\ref{del})]
\bes
\label{GeffbNf}
\bea
\Gamma_{\rm eff}(h \to \nu_k \ell_2^{\mp} U D; L) & = &
\int_{0}^{L} d L' \frac{d \Gamma_{\rm eff}(h \to \nu_k \ell_2^{\mp} U D; L')}{d L'}
  \label{GeffbNfdef} \\
  & = &
{\overline \Gamma}(h \to \nu_k \ell_2 U D)  |U_{\ell_2 N}|^2 {\Bigg \{}
\left[ ({\cal A}_1^{(k)})^2 +  ({\cal A}_2^{(k)})^2 \right]
\left[1 - \exp \left(- L \frac{\Gamma_N}{\beta_N \gamma_N} \right) \right]
\nonumber\\  &&
+ 2 {\cal A}_1^{(k)} {\cal A}_2^{(k)} {\Bigg [}
  {\bigg [} - \frac{1}{(1 + y^2)} \cos \left( 2\pi \frac{L}{L_{\rm osc}} \right) \cos (\Delta \phi)
  \mp 2  \frac{y}{(1 + y^2)^2} \cos \left( 2\pi \frac{L}{L_{\rm osc}} \right) \sin (\Delta \phi)
 \nonumber\\  &&
 \mp \frac{(1-y^2)}{(1 + y^2)^2} \sin \left( 2\pi \frac{L}{L_{\rm osc}} \right) \sin (\Delta \phi) {\bigg ]}
 \exp \left(- L \frac{\Gamma_N}{\beta_N \gamma_N} \right)
 \nonumber\\  &&
 + \left[ \frac{1}{(1 + y^2)}  \cos(\Delta \phi) \pm 2 \frac{y}{(1 + y^2)^2} \sin (\Delta \phi) \right] {\Bigg ]} {\Bigg \}}.
\label{GeffbNfexpr}
\eea \ees

\begin{figure}[htb] 
\centering\includegraphics[width=100mm]{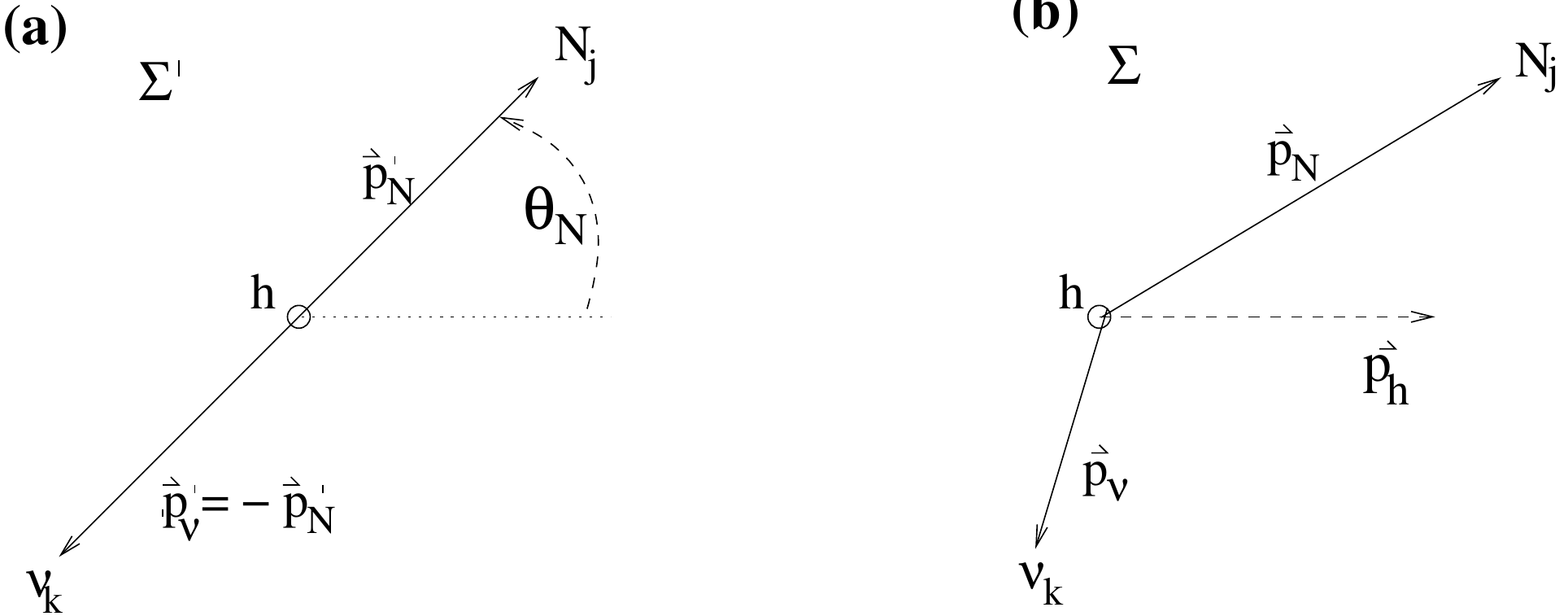}
\caption{\footnotesize (a) The 3-momenta of the produced particles in the decay $h \to N_j \nu_k$ in the Higgs rest frame ($\Sigma^{'}$); (b) the same, but in the lab frame ($\Sigma$).}
\label{FigSpS}
 \end{figure}

We note that the speed $\beta_N$ of the produced intermediate $N_j$, appearing in these formulae, is in the lab frame $\Sigma$. In the above formulas we considered $\beta_N$ to be fixed. However, in practice this is not the case. It is the speed $\beta_N^{'}$ of $N_j$ in the $h$-rest frame $\Sigma^{'}$ that is fixed
\bes
\label{kinSigp}
\bea
E_N^{'} & = & \frac{M_h^2 + M_N^2}{2 M_h}, \; |{\vec p'}_N|= \frac{1}{2} M_h \left( 1 - \frac{M_N^2}{M_h^2} \right),
\label{ENppNp} \\
\beta_N^{'} \gamma^{'}_N &=& \sqrt{(E_N^{'}/M_N)^2 - 1} = \frac{M_h^2-M_N^2}{ 2 M_h M_N}.
\label{bNgNp} \eea \ees
On the other hand, it is expected that the velocity (the speed and the direction) ${\vec \beta}_h$ of the produced Higgs particles in the lab frame $\Sigma$ is approximately fixed. We can regard this direction as the $z$-axis direction in the $N_j$-rest frame $\Sigma^{'}$, i.e., ${\hat \beta}_h \equiv {\hat z}^{'}$.
When we go from the $h$-rest frame ($\Sigma^{'}$) to the lab frame ($\Sigma$), the corresponding quantities $E_N$ and $\beta_N \gamma_N$ there will depend on the angle $\theta_N$ between the ${\vec p'}_N$ and ${\hat z}^{'}$ ($={\hat \beta}_h$) (cf.~Fig.~\ref{FigSpS})
\bes
\label{kinSig}
\bea
E_N & = & \gamma_h (E^{'}_N + \cos \theta_N \beta_h |{\vec p'}_N|),
\label{EN} \\
\beta_N \gamma_N &=& \sqrt{\left( \frac{E_N}{M_N} \right)^2 - 1} =
\sqrt{ \gamma_h^2 \left( \frac{E^{'}_N + \cos \theta_N \beta_h |{\vec p'}_N|}{M_N} \right)^2 - 1} = \beta_N \gamma_N(\theta_N),
\label{bNgN} \eea \ees
where $E^{'}_N$ and $|{\vec p'}_N|$ are the constants given in Eq.~(\ref{ENppNp}). Similar considerations were made, e.g., in Ref.~\cite{GCCSK2017} (where the lab frame was denoted by $\Sigma^{''}$).

On the other hand, we see that the effective decay width (\ref{GeffbNfexpr}) has no dependence on the direction ${\hat p'}_N$ of $N_j$ in its rest frame (i.e., on $\theta_N$), with the exception of the Lorentz factor $1/(\gamma_N \beta_N)$ [cf.~Eq.~(\ref{bNgN})]  appearing in the exponent $\exp(- L \Gamma_N/(\gamma_N \beta_N))$ (in two places) and in $(2 \pi/L_{\rm osc})$, cf. Eq.~(\ref{Losc}). We recall that this factor originates from the decay probabilities $d P_N(L; \theta_N)/d L$. We can then write
\be
\Gamma_{\rm eff}(h \to \nu_k \ell_2^{\mp} U D; L) =
\int d \Omega_{{\hat p'}_N} \frac{\Gamma_{\rm eff}(h \to \nu_k \ell_2^{\mp} U D; L; \theta_N)}{ d \Omega_{{\hat p'}_N} },
\label{dGeffdOm} \ee
where we use $d \Omega_{{\hat p'}_N} = 2 \pi d(\cos \theta_N)$. With this $\theta_N$-dependence, the quantity of Eq.~(\ref{dGeff4}), in the scenario Eq.~(\ref{dPN}) [cf.~also Eq.~(\ref{dPNj})] can be written as
\bea
\lefteqn{
\frac{d \Gamma_{\rm eff}(h \to \nu_k \ell_2^{\mp} U D; L; \theta_N)}{2 \pi d \cos \theta_N} =
\frac{d {\overline \Gamma}(h \to \nu_k \ell_2 U D)}{2 \pi d \cos \theta_N}
dP_N(L; \theta_N)
{\bigg \{}
\frac{\Gamma_N}{\Gamma_{N_1}} ({\cal A}^{(k)}_1)^2  |U_{\ell_2 N_1}|^2  +
\frac{\Gamma_N}{\Gamma_{N_2}} ({\cal A}^{(k)}_2)^2  |U_{\ell_2 N_2}|^2 +
}
\nonumber\\ &&
+
2 {\cal A}^{(k)}_1 {\cal A}^{(k)}_2 |U_{\ell_2 N_1}| |U_{\ell_2 N_2}|
\left[ \delta(y) \cos\left( 2 \pi \frac{L}{L_{\rm osc}} \mp \Delta \phi \right)
  + \frac{ \eta(y)}{y} \sin\left( 2 \pi \frac{L}{L_{\rm osc}} \pm \Delta \phi \right) \right]  {\bigg \}},
\label{dGeff4thetaN} \eea
where now $2 \pi/L_{\rm osc}$ and the differential decay probability $dP_N$ are $\theta_N$-dependent
\bes
\label{LoscdPN}
\bea
\frac{2 \pi}{L_{\rm osc}(\theta_N)} &=& y \frac{\Gamma_N}{\beta_N \gamma_N(\theta_N)},
  \label{LoscthetaN} \\
 dP_N(L; \theta_N)  &=& \frac{\Gamma_N}{\beta_N \gamma_N (\theta_N)} \exp \left( - L \frac{\Gamma_N}{\beta_N \gamma_N (\theta_N)} \right),
\label{dpNthetaN} \eea \ees
because of the $\theta_N$-dependence of the product $\beta_N \gamma_N$, cf.~Eq.~(\ref{bNgN}). However, the expression $d {\overline \Gamma}(h \to \nu_k \ell_2 U D; \theta_N)$ is $\theta_N$-independent, because Higgs $h$ is a scalar [i.e., in Eq.~(\ref{fact}) the decay width $d {\overline \Gamma}(h \to \nu_k N_j)$ is $\theta_N$-independent]. As a consequence, the first factor on the right-hand side of Eq.~(\ref{dGeff4thetaN}) becomes
\be
\frac{d {\overline \Gamma}(h \to \nu_k \ell_2 U D)}{2 \pi d \cos \theta_N}
= \frac{1}{4 \pi} {\overline \Gamma}(h \to \nu_k \ell_2 U D),
\label{dbarGrepl} \ee
where ${\overline \Gamma}(h \to \nu_k \ell_2 U D)$ is given in Eqs.~(\ref{factbarGNext}).
This then implies that the general expression for the effective decay width is obtained from the expression (\ref{GeffbNf}) with the replacement of the $\theta_N$-dependent exponents and $2 \pi/L_{\rm osc}(\theta_N)$ there by integration over $2 \pi d \cos \theta_N$ and dividing by the overall factor $4 \pi$
\bes
\label{Geff}
\bea
\Gamma_{\rm eff}(h \to \nu_k \ell_2^{\mp} U D; L) & = &
\int_{0}^{L} d L' \int d \Omega_{{\hat p'}_N} \frac{d^2 \Gamma_{\rm eff}(h \to \nu_k \ell_2^{\mp} U D; L')}{d L' d \Omega_{{\hat p'}_N}}
  \label{Geffdef} \\
  & = &
        {\overline \Gamma}(h \to \nu_k \ell_2 U D)  \; |U_{\ell_2 N}|^2 \times
        \frac{1}{2} \int_{\cos \theta_N=-1}^{+1} d \cos \theta_N
 \nonumber\\ && \times
{\Bigg \{}
\left[ ({\cal A}_1^{(k)})^2 +  ({\cal A}_2^{(k)})^2 \right]
\left[1 -   \exp \left(- L \frac{\Gamma_N}{\beta_N \gamma_N(\theta_N)} \right) \right]
\nonumber\\  &&
+ 2 {\cal A}_1^{(k)} {\cal A}_2^{(k)} {\Bigg [}
  {\bigg [} - \frac{1}{(1 + y^2)} \cos \left( 2\pi \frac{L}{L_{\rm osc}(\theta_N)} \right) \cos (\Delta \phi)
  \mp 2  \frac{y}{(1 + y^2)^2} \cos \left( 2\pi \frac{L}{L_{\rm osc}(\theta_N)} \right) \sin (\Delta \phi)
 \nonumber\\  &&
 \mp \frac{(1-y^2)}{(1 + y^2)^2} \sin \left( 2\pi \frac{L}{L_{\rm osc}(\theta_N)} \right) \sin (\Delta \phi) {\bigg ]}
  \exp \left(- L \frac{\Gamma_N}{\beta_N \gamma_N(\theta_N)} \right)
 \nonumber\\  &&
 + \left[ \frac{1}{(1 + y^2)}  \cos(\Delta \phi) \pm 2 \frac{y}{(1 + y^2)^2} \sin (\Delta \phi) \right] {\Bigg ]} {\Bigg \}},
\label{Geffexpr}
\eea \ees
where we use for the Lorentz factor product $\beta_N \gamma_N (\theta_N)$, which appears in $2 \pi/L_{\rm osc}(\theta_N)$ and in the exponents, the expression (\ref{kinSig}) [in conjunction with Eqs.~(\ref{kinSigp})].

\begingroup \color{black}
We point out that the index $k$ ($k=1,2,3$) in Eqs.~(\ref{Geff}) refers to the light neutrino mass eigenstate $\nu_k$  ($m_{\nu_k} < 1$ eV), and that from the detection point of view the sum of the decay widths (over $k=1,2,3$) can be measured, but not the separate decay widths because the light neutrino is not directly detected. Nonetheless, if we assume that in the Yukawa couplings ${\cal A}_j^{(k)}$, Eq.~(\ref{Aj}), the term proportional to $G_{\nu_{\tau}}$ is the dominant one [cf.~Eq.~(\ref{YuNtau})], then $k=3$ is expected to be the dominant term because among the PMNS elements $U_{\tau \nu_k}$ the element $U_{\tau \nu_3}$ is the dominant. In the numerical analysis in the next Section we will take $\ell_2=\tau$ and will assume that $k=3$.

The obtained final expression Eq.~(\ref{Geffexpr}) has three terms with opposite signs for the two channels $h \to \nu_k \ell_2^{\mp} U D$, all three proportional to $\sin(\Delta \phi)$ where we recall that $\Delta \phi$ is the difference of the phases of the heavy-light mixing parameters $U_{\ell_2 N_j}$ ($j=1,2$), cf.~Eq.~(\ref{delphi}). This now implies that in general we obtain for such processes a nonzero CP violation asymmetry
\be
A_{\rm CP} = \frac{ \Gamma(h \to \nu_k \ell_2^- U {\bar D}) - \Gamma(h \to \nu_k \ell_2^+ {\bar U} D) } { \Gamma(h \to \nu_k \ell_2^- U {\bar D}) + \Gamma(h \to \nu_k \ell_2^+ {\bar U} D) },
\label{ACP} \ee
and that this asymmetry is proportional to $\sin(\Delta \phi)$. We point out that the neutrinos were considered to be Majorana throughout this work. If, on the other hand, the neutrinos were Dirac, then we would have in the first vertex of the considered rare decays, instead of the (real) Yukawa couplings ${\cal A}_j^{(k)}$ [cf.~Eqs.~(\ref{YuN})-(\ref{Aj})], different and complex Yukawa couplings. Namely, it can be explicitly checked that for the decay $h \to {\bar \nu}_k \ell_2^{-} \pi^+$ (or: $h \to {\bar \nu}_k \ell_2^{-} U {\bar D}$) we would obtain for the corresponding amplitude the same result as ${\cal T}_{-}$ Eq.~(\ref{Tm}), but with the replacement ${\cal A}_j^{(k)} \mapsto (1/2) G_{\nu_{\ell}} U_{\ell N_j}^{\ast} U_{\ell \nu_k}$ (sum over $\ell=e, \mu, \tau$); and for the charge-conjugate channel $h \to \nu_k \ell_2^{+} \pi^-$  the decay amplitude ${\cal T}_{+}$ Eq.~(\ref{Tp}) with the replacement ${\cal A}_j^{(k)} \mapsto (1/2) G_{\nu_{\ell}} U_{\ell N_j} U_{\ell \nu_k}^{\ast}$. This implies that the effective phases in the amplitudes ${\cal T}_{\mp}$ in the Dirac case are not any more $\pm \phi_j$ [the phases of $U_{\ell_2 N_j}$and $U_{\ell_2 N_j}^{\ast}$ Eq.~(\ref{delphi})], but rather the phases $\pm (\phi_j - \theta_{j k})$ where $\theta_{j k}$ is the phase of the complex Yukawa coupling $\sum_{\ell} (1/2) G_{\nu_{\ell}} U_{\ell N_j} U_{\ell \nu_k}^{\ast}$
\be
\sum_{\ell=e,\mu,\tau}  G_{\nu_{\ell}} U_{\ell N_j} U_{\ell \nu_k}^{\ast} =
{\Big |} \sum_{\ell=e,\mu,\tau} G_{\nu_{\ell}} U_{\ell N_j} U_{\ell \nu_k}^{\ast} {\Big |} \;  e^{i \theta_{j k}}.
\label{thetajk} \ee
If, however, in this sum only the $\ell = \tau$ term is dominant [cf.~Eq.~(\ref{YuNtau})], and we take for $\ell_2=\tau$ (i.e., the case taken in the next Section), then $\theta_{j k} = + \phi_j - \eta_k$ where $\eta_k$ is the phase of the PMNS element $U_{\tau \nu_k}$. In such a case, the phase of ${\cal T}_{\mp}$ (which was in the Majorana case $\pm \phi_j$) reduces to $\pm (\phi_j - \theta_{j k}) = \pm \eta_k$, i.e., this phase is now independent of $j=1,2$. In this case, $\Delta \phi \equiv \phi_2 - \phi_1$, Eq.~(\ref{delphi}), reduces to zero, and we have $A_{CP} = 0$. For this reason, we expect that in the case of the Dirac neutrinos the CP violation asymmetry is either zero or it is suppressed with respect to the Majorana neutrino case.
\endgroup

\section{Analysis and results at ILC}
\label{sec:ANARES}

\begin{figure}[htb] 
\includegraphics[width=85mm]{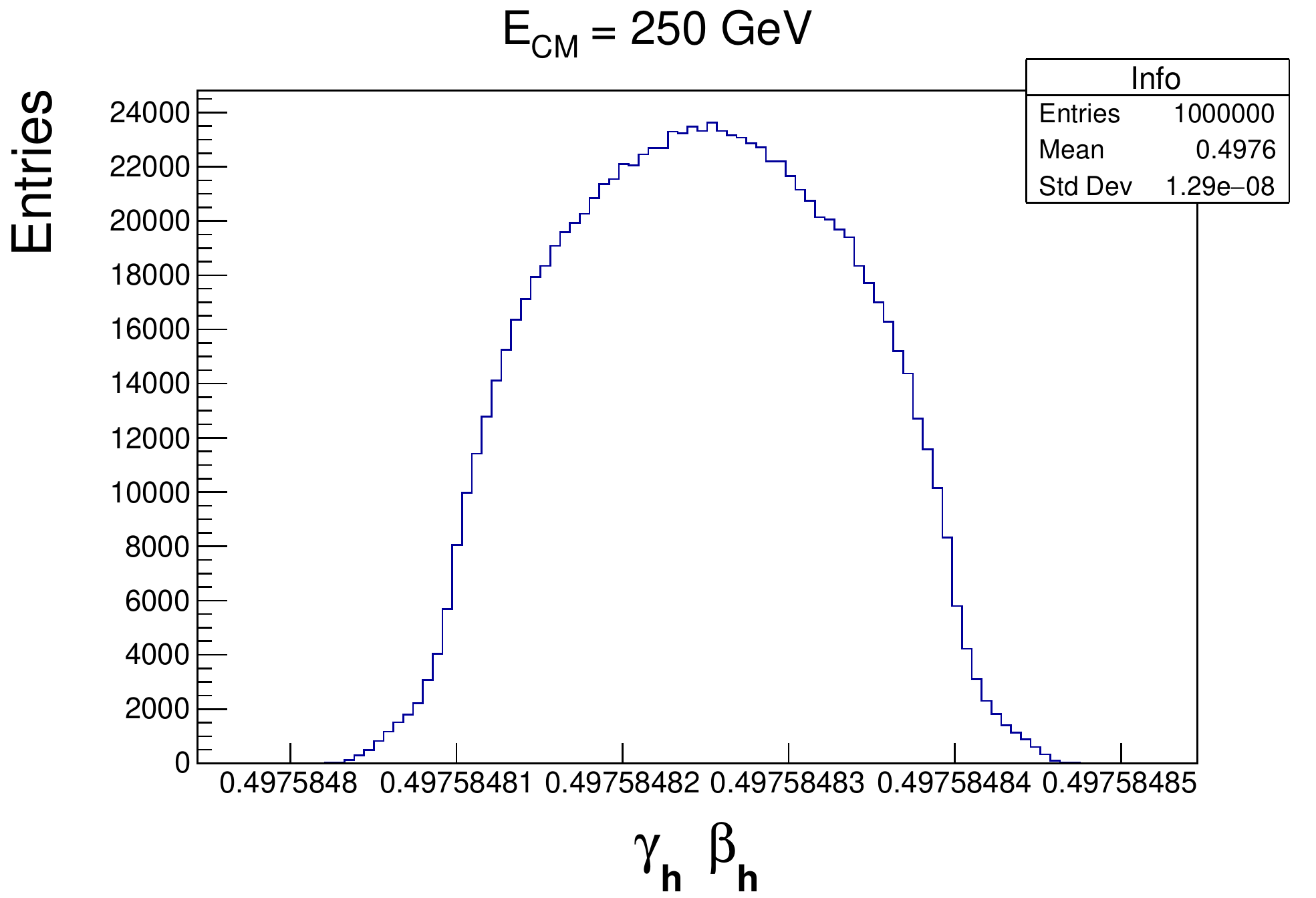}
\includegraphics[width=85mm]{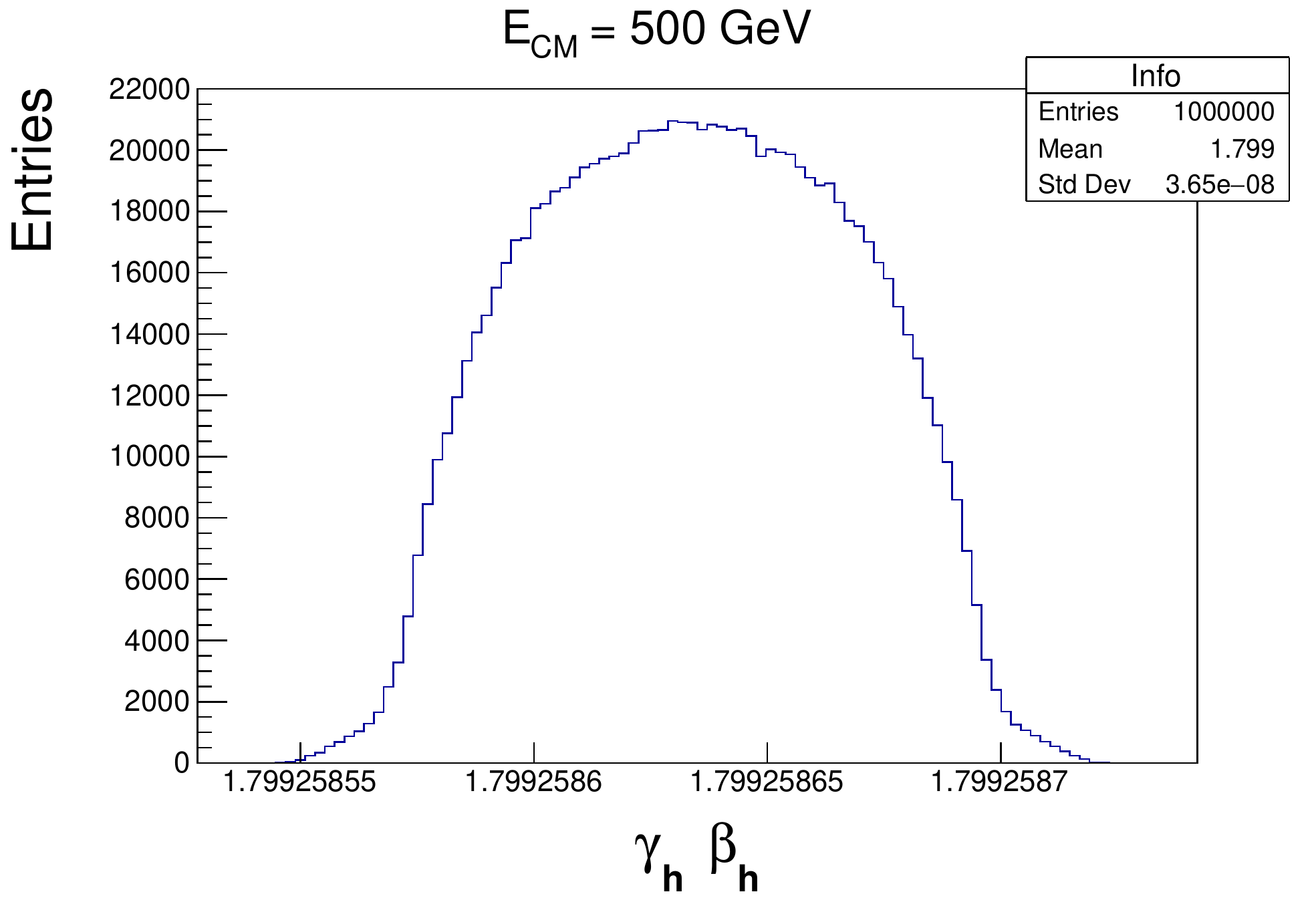}
\caption{The $\gamma_h \beta_h$ distributions for the produced Higgs bosons. Left panel: $\sqrt{s}=250$GeV; right panel: $\sqrt{s}=500$GeV.}
\label{Fig.gambet}
 \end{figure}

A proper evaluation of the heavy sterile neutrino (HSN) energy in Eq.~(\ref{EN}) requires a realistic distribution of $\beta_h \gamma_h$. This distribution is obtained by simulating $10^6$ Higgs bosons produced in a $e^+$$e^-$ collider, using \textsc{MadGraph5\_aMC@NLO}~\cite{Alwall:2014hca}, \textsc{Pythia8}~\cite{Sjostrand:2007gs} and \textsc{Delphes}~\cite{deFavereau:2013fsa}, for the ILC conditions for $\sqrt{s}=250$ GeV and $\sqrt{s}=500$ GeV. The obtained results are presented in Fig.~\ref{Fig.gambet}; they give us $\beta_h \gamma_h \approx 0.4976$ and $1.799$ for $\sqrt{s}=250$ GeV and $\sqrt{s}=500$ GeV, respectively.

The integrated luminosities (based on 20 years of operation) expected at ILC are $1150 \ {\rm fb}^{-1}$ and $1600 \ {\rm fb}^{-1}$ for $\sqrt{s}=250$ GeV and $\sqrt{s}=500$ GeV, respectively (see Ref.~\cite{Asner:2013psa}).
The Higgs production cross-sections are $\sigma_{h} \approx 620 \ {\rm fb}$
and $\sigma_{h} \approx 510 \ {\rm fb}$ for $\sqrt{s} =250$ GeV and $\sqrt{s}=500$ GeV, respectively (see Ref.~\cite{Yamamoto:2021kig}). Therefore, the expected numbers of produced Higgs bosons ($N_h$), after 20 years of operation, are as given in Table I.
\begin{table}[htb]
\begin{center}
\begin{tabular}{c|c|c|c|c}
Energy & $\sigma_{h}$ $({\rm fb})$ & $\int \cal{L}$ $dt$ (${\rm fb}^{-1}$) & efficiency & $N_{h}$\\
\hline
$\sqrt{s}=250$GeV & $620$ & $1150$ & 0.3 & $2.1 \times 10^{5}$ \\
$\sqrt{s}=250$GeV & $620$ & $1150$ & 0.5 & $3.6 \times 10^{5}$ \\
$\sqrt{s}=500$GeV & $510$ & $1600$  & 0.3 & $2.5 \times 10^{5}$  \\
$\sqrt{s}=500$GeV & $510$ & $1600$  & 0.5 & $4.1 \times 10^{5}$  \\
\end{tabular}
\label{Nhtab}
\caption{Expected number of produced Higgs bosons.}
\end{center}
\end{table}

\begin{figure}[htb] 
\includegraphics[width=85mm]{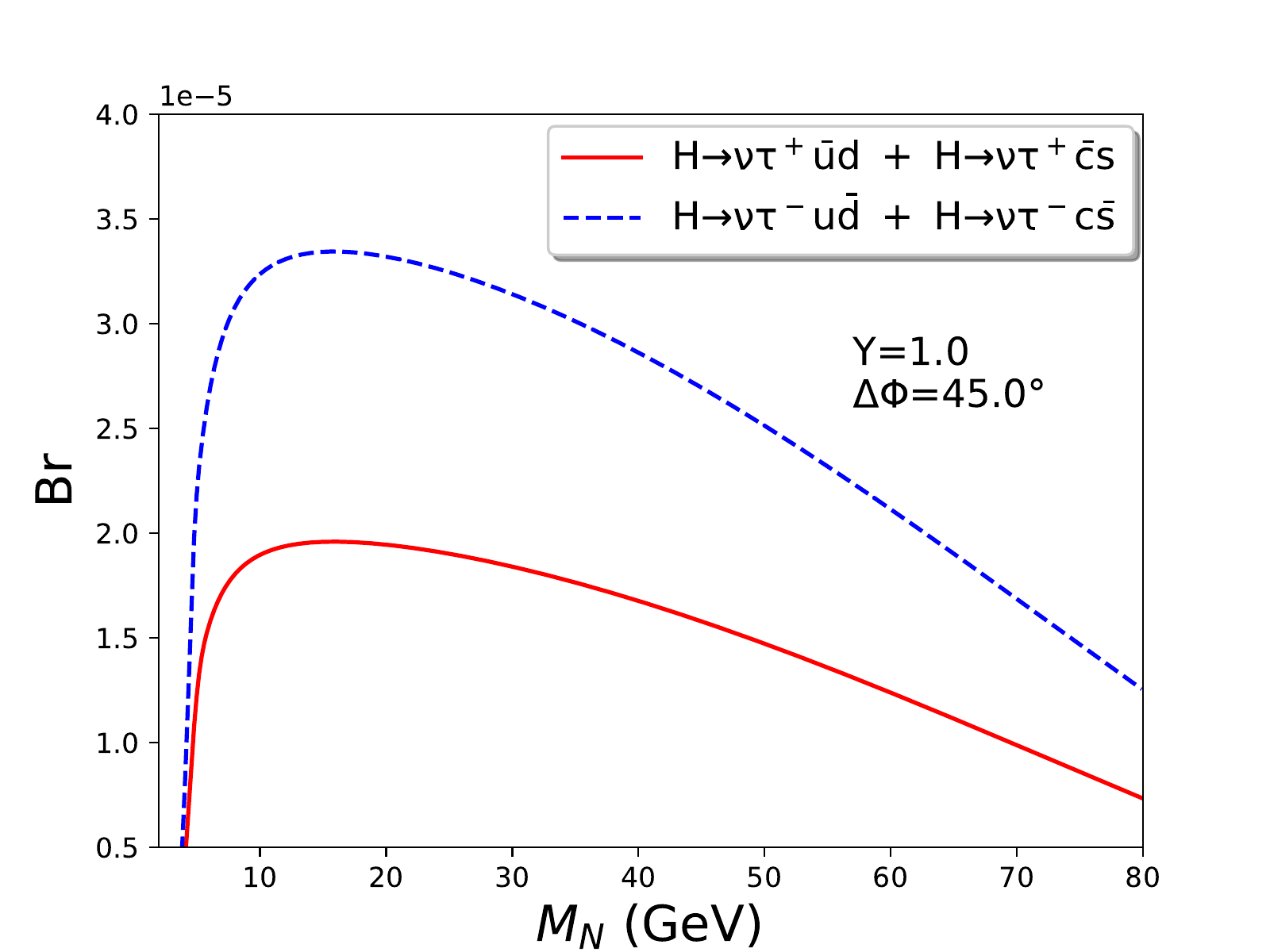}
\includegraphics[width=85mm]{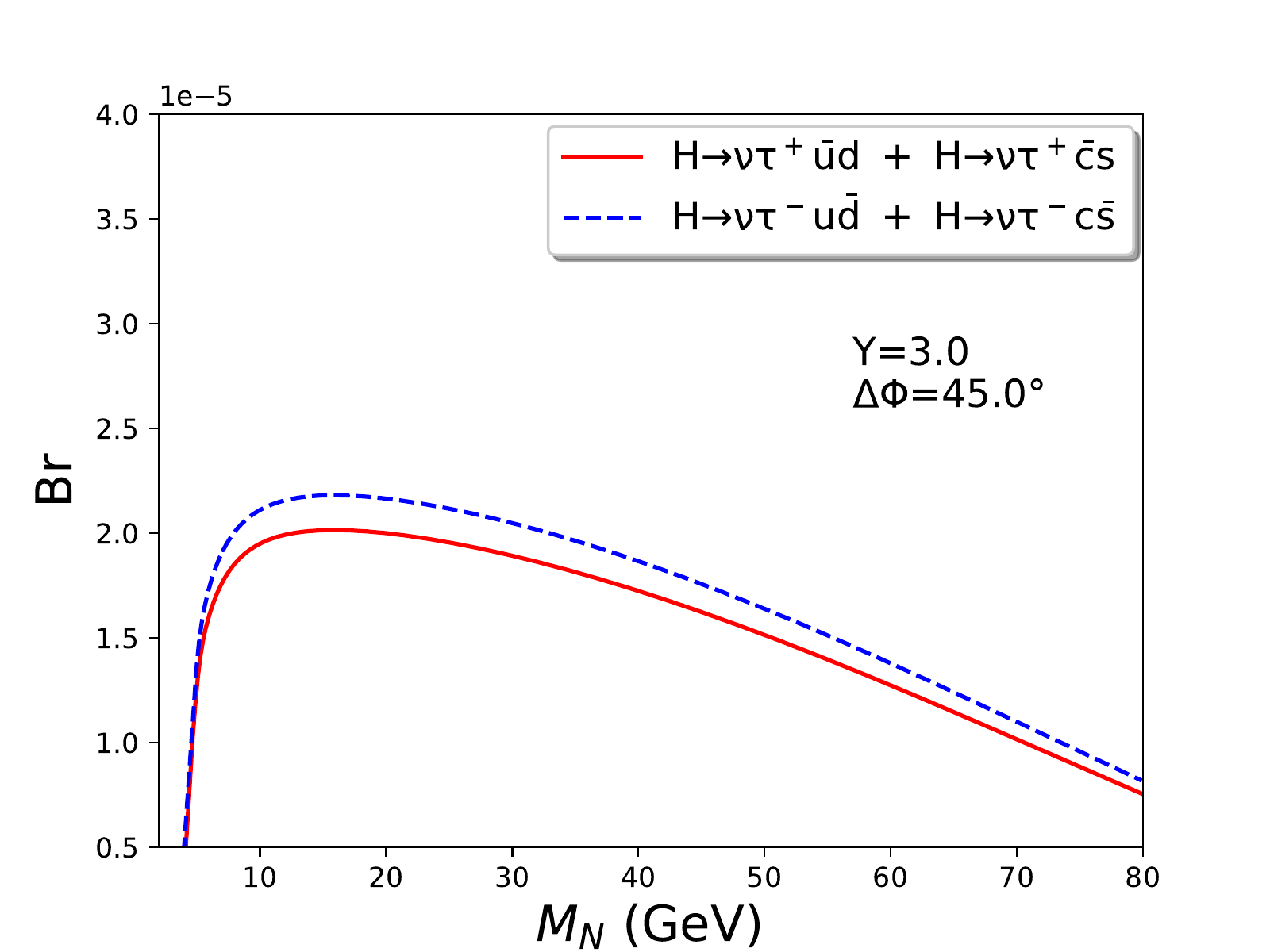}
\caption{ Higgs boson Branching ratio as a function of HSN mass for different values of $y \equiv \Delta M_N/\Gamma_N$ ($=1.0$, $3.0$) and $\Delta \phi=45^o$. Here we used $|U_{\tau N}|^2=10^{-5}$, $L=1$ m, and $\sqrt{s}=250$ GeV. The Yukawa couplings ware set equal to ${\cal A}_1^{(k)}={\cal A}_2^{(k)}=1.18 \times 10^{-4}$ (corresponding to $G_{\nu_{\tau}} \approx 0.06$, cf.~discussion in the text). Here ${\rm Br(H \to \nu_k \tau U D)}$ include the ${\rm Br(H \to \nu_k \tau c s)}$ and ${\rm Br(H \to \nu_k \tau u d)}$ channels.}
\label{Fig.ANA250-1}
\end{figure}

\begin{figure}[htb] 
\includegraphics[width=85mm]{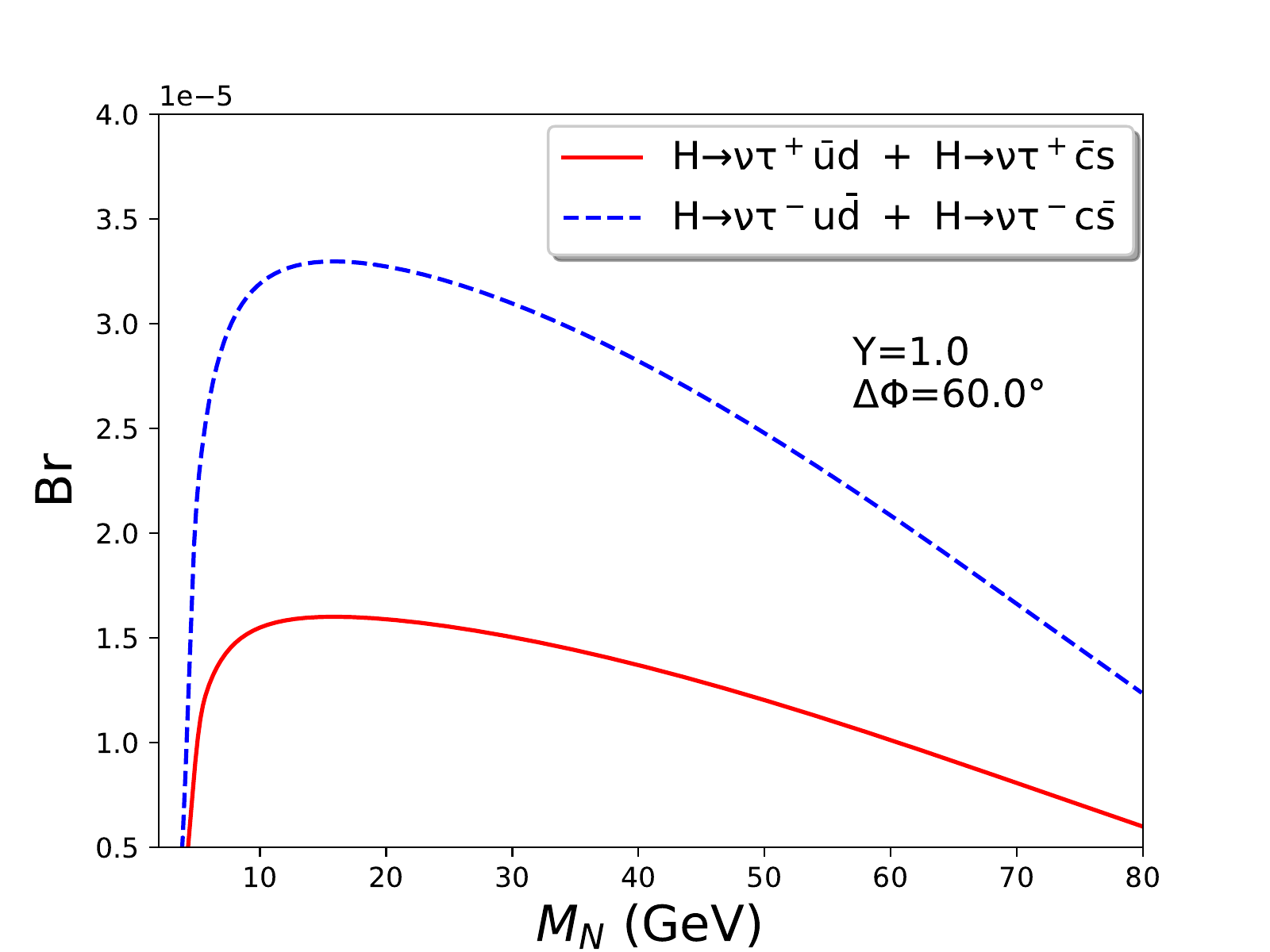}
\includegraphics[width=85mm]{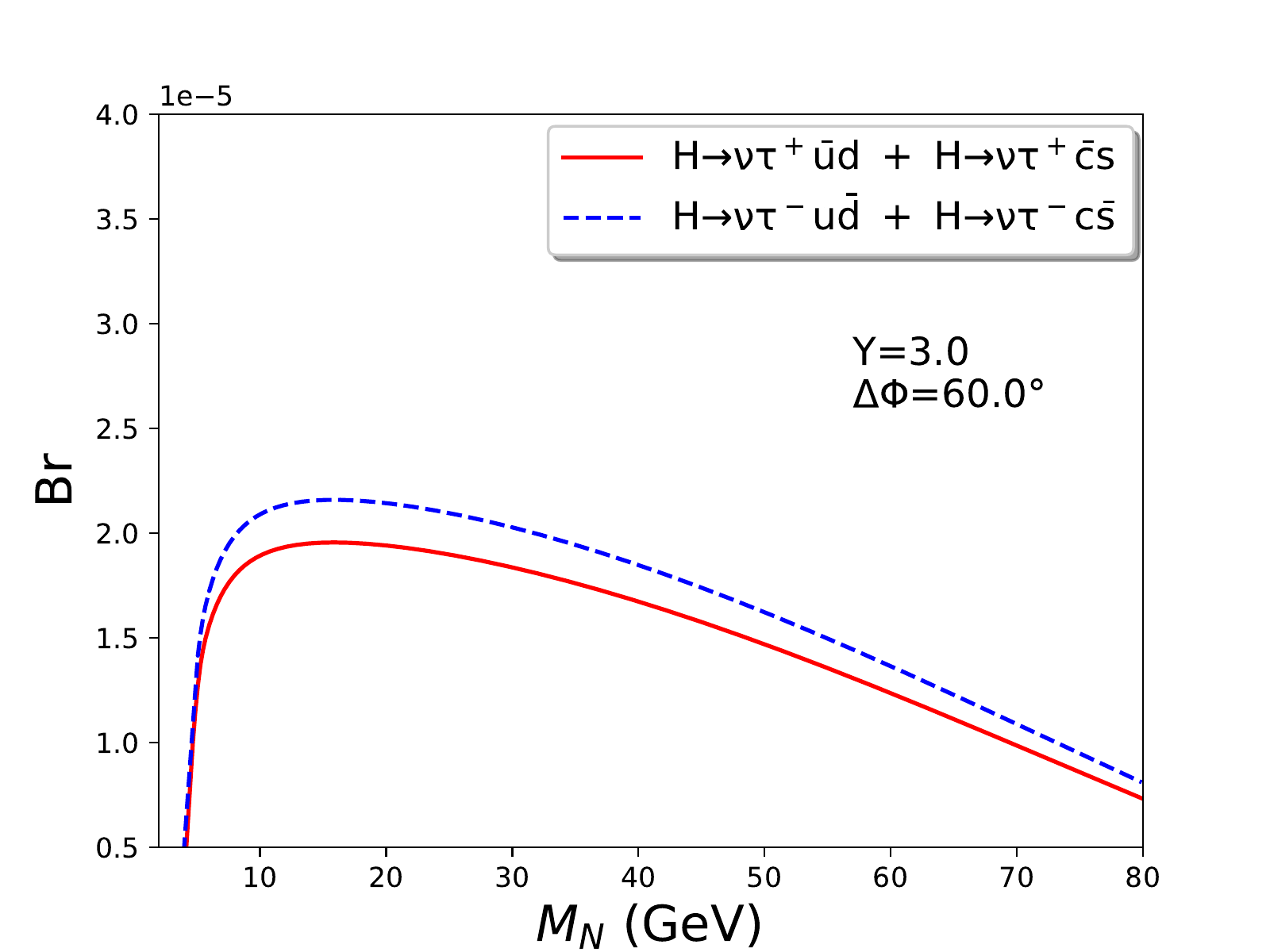}
\caption{ Higgs boson Branching ratio as a function of HSN mass for differents values of $y$ and $\Delta \phi=60^o$. Here we used $|U_{\tau N}|^2=10^{-5}$, $L=1$ m, ${\cal A}_1^{(k)}={\cal A}_2^{(k)}=1.18 \times 10^{-4}$ and $\sqrt{s}=250$ GeV .}
\label{Fig.ANA250-2}
\end{figure}

\begin{figure}[htb] 
\includegraphics[width=85mm]{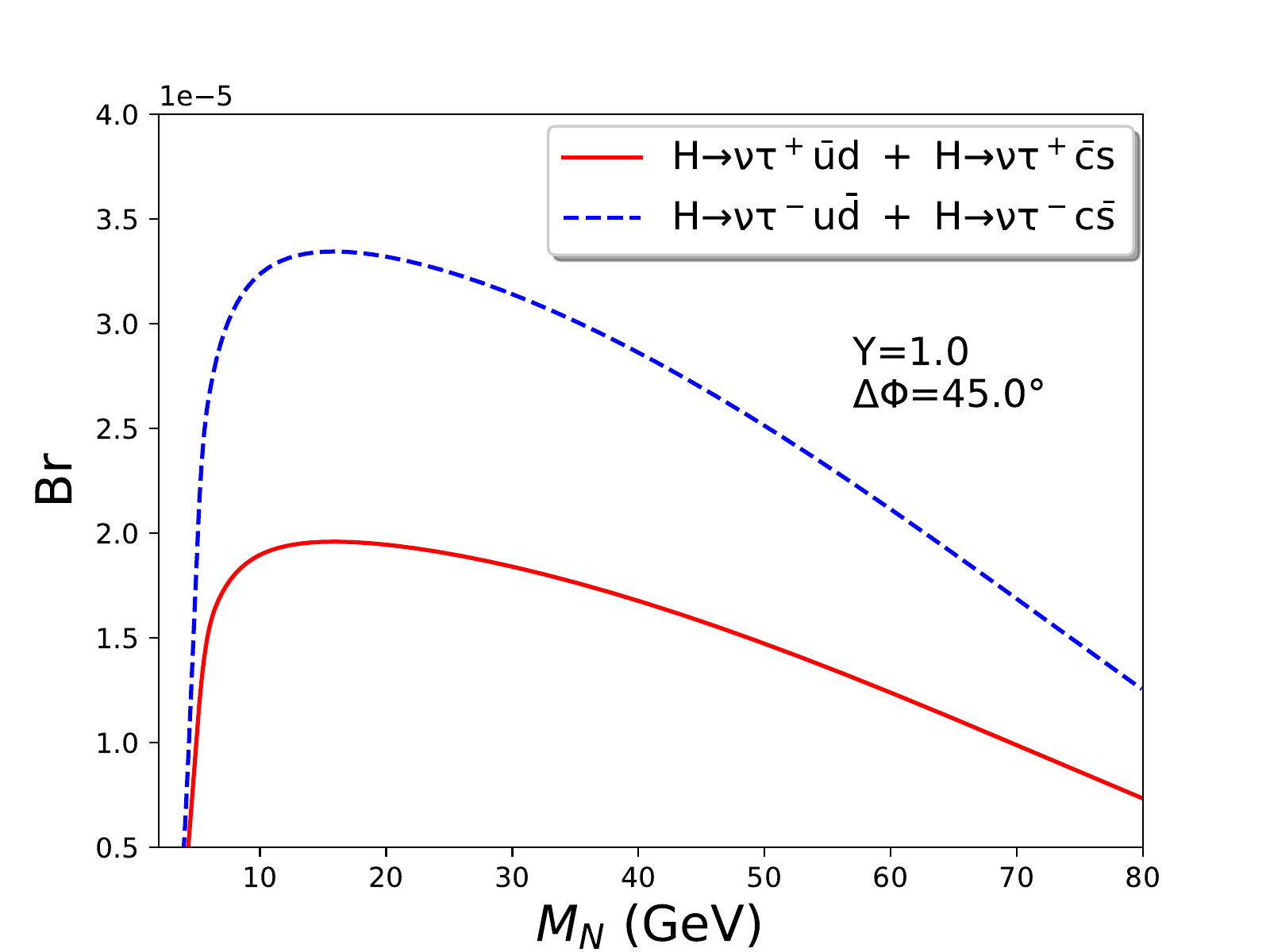}
\includegraphics[width=85mm]{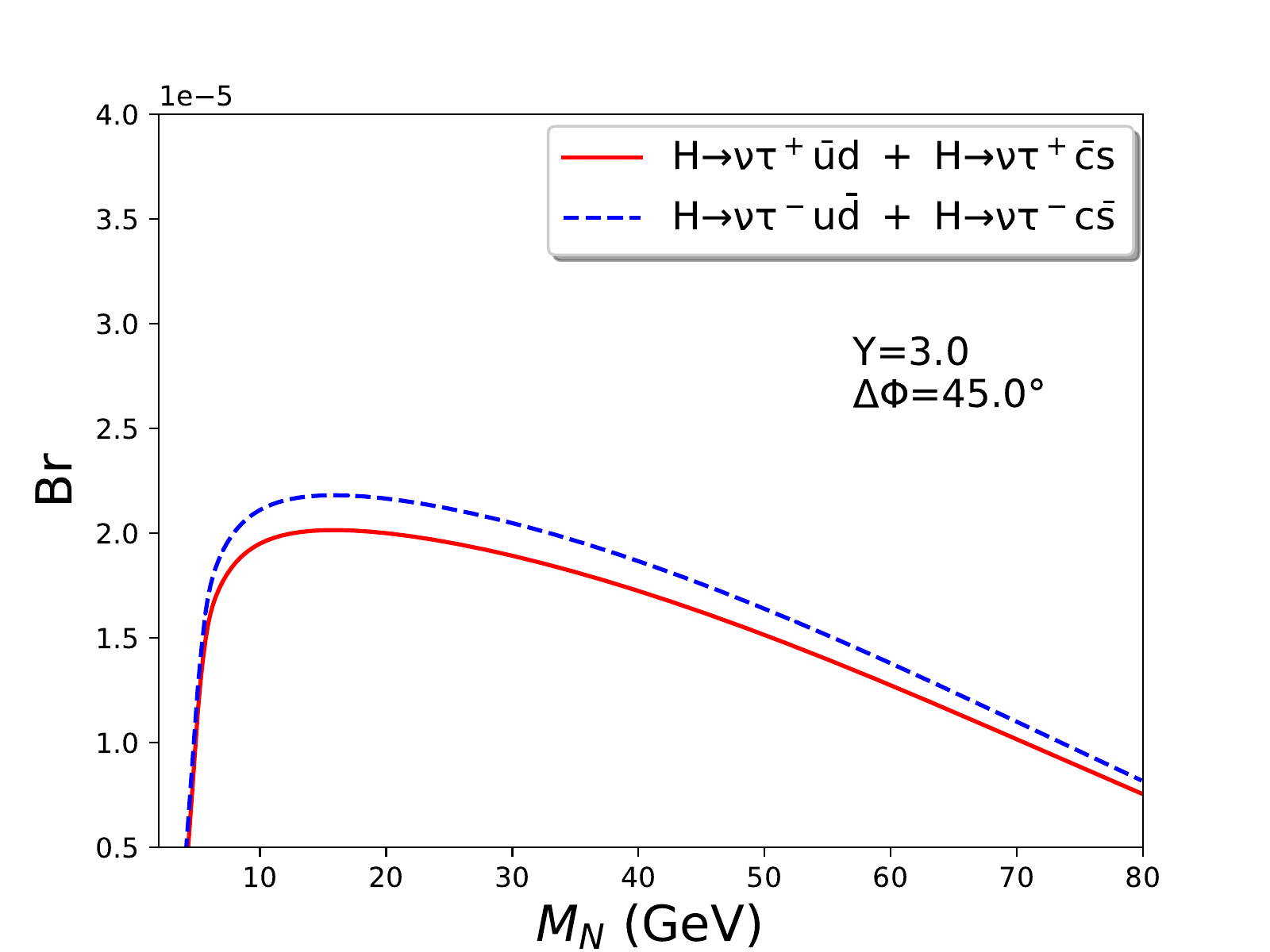}
\caption{ Higgs boson Branching ratio as a function of HSN mass for differents values of $y$ and $\Delta \phi = 45^o$. Here we used $|U_{\tau N}|^2=10^{-5}$, $L=1$ m, ${\cal A}_1^{(k)}={\cal A}_2^{(k)}=1.18 \times 10^{-4}$ and $\sqrt{s}=500$ GeV .}
\label{Fig.ANA500-1}
\end{figure}

\begin{figure}[htb] 
\includegraphics[width=85mm]{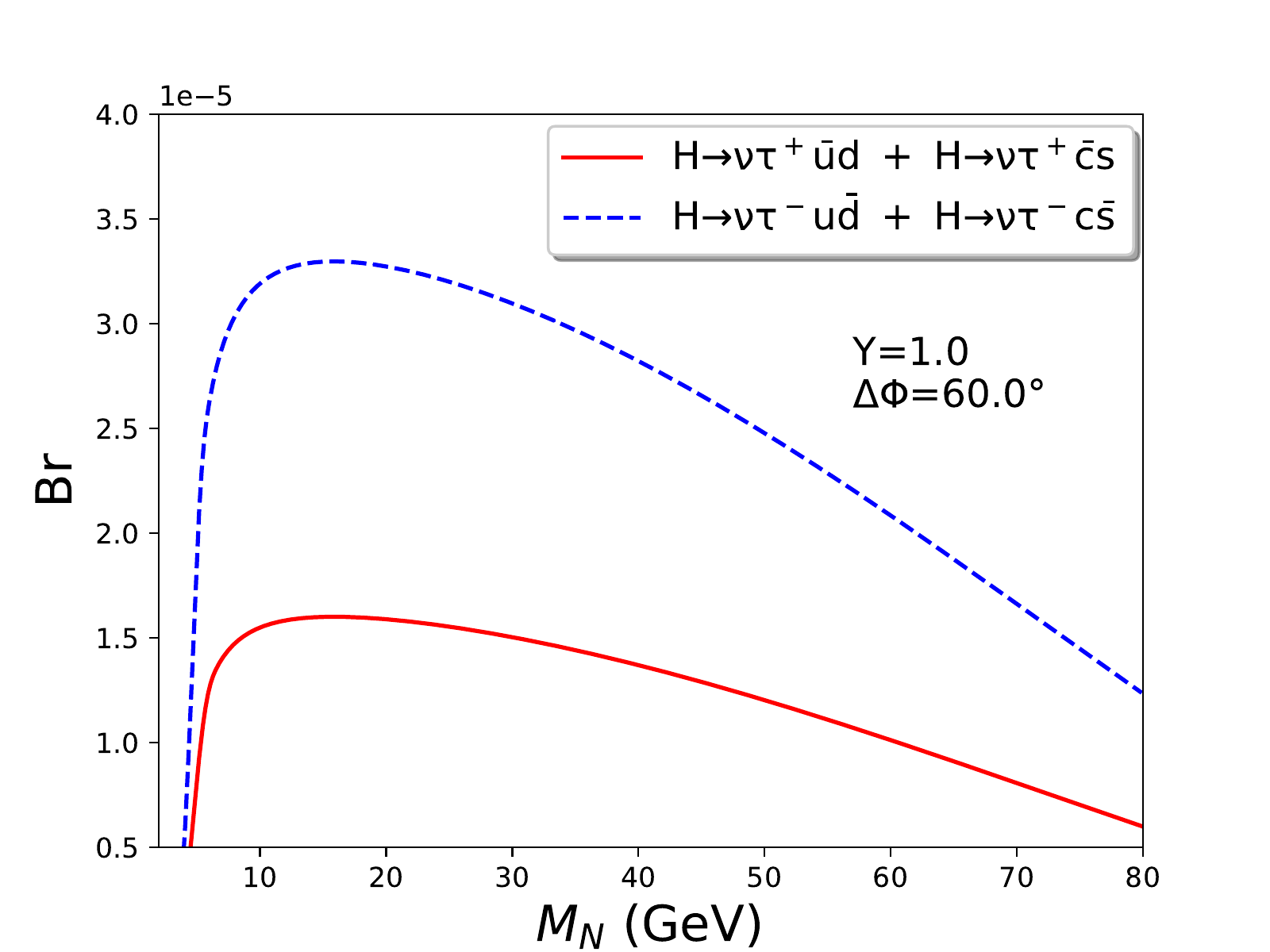}
\includegraphics[width=85mm]{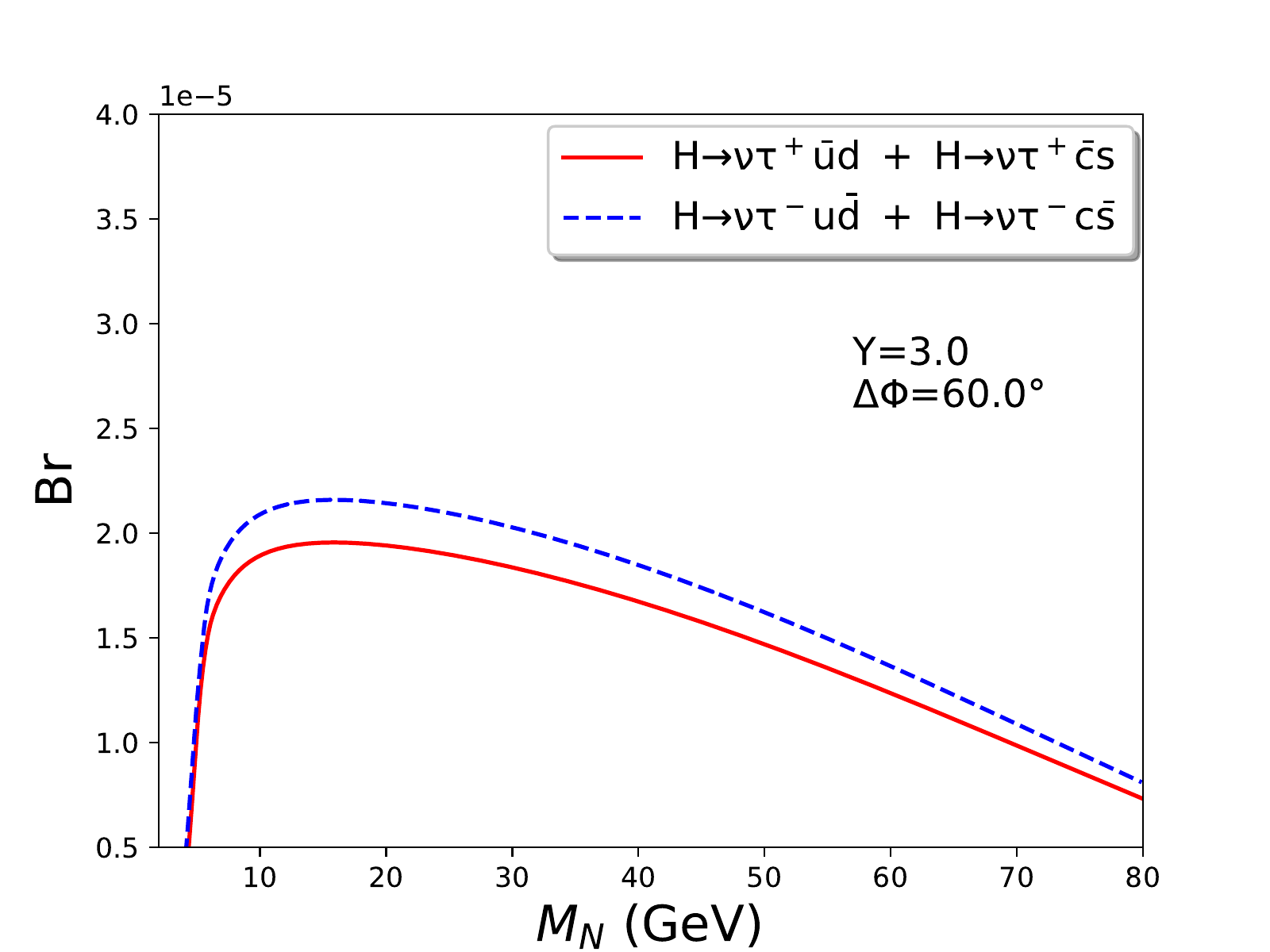}
\caption{ Higgs boson Branching ratio as a function of HSN mass for differents values of $y$ and $\Delta \phi=60^o$. Here we used $|U_{\tau N}|^2=10^{-5}$, $L=1$ m, ${\cal A}_1^{(k)}={\cal A}_2^{(k)}=1.18 \times 10^{-4}$ and $\sqrt{s}=500$ GeV .}
\label{Fig.ANA500-2}
\end{figure}

The Higgs boson Branching Ratio is ${\rm Br(h)} \equiv {\rm Br(h \to \nu \tau q \bar{q})}/\Gamma({\rm h \to all})$, were $\Gamma({\rm h \to all}) = 3.2\times 10^{-3}$ GeV is the total Higgs boson decay width \cite{CMS:2019ekd}. In Figs.~\ref{Fig.ANA250-1}-\ref{Fig.ANA500-2} we present the results for ${\rm Br(h)}$ as a function of the HSN mass, for various scenarios.


Inspection of Figs.~\ref{Fig.ANA250-1}-\ref{Fig.ANA500-2}, in conjunction with the data in Table I, indicates that if the efficiency $0.5$ is achieved, the number of detected rare Higgs decays $h \to \nu_k \tau^{\mp} U D$ is $\sim 10^1$ for the mentioned input parameters, in the heavy neutrino mass range $10 \ {\rm GeV} < M_N < 50 \ {\rm GeV}$. When $y (\equiv \Delta M_N/\Gamma_N) \approx 1$, then $\sim 10^1$ is also roughly the number of the difference of the events with $h \to \nu_k \tau^{-} U D$ and $h \to \nu_k \tau^{+} U D$ provided $\Delta \phi$ is in the range between $45^o$ and $60^o$. This would make it possible to detect the CP violation \textcolor{black}{asymmetry $A_{\rm CP}$ Eq.~(\ref{ACP}) with $\ell_2=\tau$.}

The value of the heavy-light mixing parameter [cf.~Eq.~(\ref{mix})] was taken to be $|U_{\tau N}|^2 = 10^{-5}$ which is the present upper bound \cite{DELPHI:1996qcc} in the mentioned mass range of $M_N$.\footnote{If we considered that the produced charged lepton $\ell^{\pm}$ at the second vertex is muon of electron, the heavy-light mixing parameter $|U_{\ell N}|^2$ would have a significantly more restrictive upper bounds in the considered mass range.}  On the other hand, the values of the heavy-light neutrino Yukawa couplings ${\cal A}_j^{(k)}$, cf.~Eqs.~(\ref{YuN})-(\ref{Aj}), were taken to be ${\cal A}_1^{(k)}={\cal A}_2^{(k)}=1.18 \times 10^{-4}$. When taking into account that the highest value of the PMNS matrix for $\tau$ is $U_{\tau \nu_k} \approx 0.65$ ($k=3$), and that $U_{\tau N_j} \approx 10^{-2.5}$, this implies [cf.~Eq.~(\ref{Aj})] the value $G_{\nu_{\tau}} \approx 5.7 \times 10^{-2}$ which is somewhat higher than the expectation (estimate) Eq.~(\ref{Gnutauest}). If we halve this value (to $G_{\nu_{\tau}} \approx 3 \times 10^{-2}$), then the mentioned numbers of events of the rare Higgs decays get reduced by a factor of $4$, and this would bring us roughly to the limit of detectability of such events.

\section{Conclusions}
\label{sec:concl}

We considered rare decays of Higgs boson $h$ mediated by two almost degenerate heavy on-shell Majorana neutrinos $N_j$, first in the case when the heavy Majorana neutrinos decay to pion [$\Gamma_{\mp} = \Gamma(h \to \nu_k N_j \to \nu_k \ell_2^{\pm} \pi^{\mp})$, cf.~Fig.~\ref{Fighdec}], and then in the case of the decay when the pion is replaced by the open quark channels: $\Gamma_{+}= \Gamma(h \to \nu_k N_j \to \nu_k \ell_2^{+} {\bar U} D)$ and $\Gamma_{-}=\Gamma(h \to \nu_k N_j \to \nu_k \ell_2^{-} U {\bar D})$. Here $\nu_k$ is the active light neutrino produced at the first vertex, $\ell_2^{\pm}=\ell^{\pm}$ is the charged lepton produced at the second vertex, and $U = u, c$ and $D=d, s$ are the open-channel quarks produced at the second vertex. In Sec.~\ref{sec:Gnoosc} we derived the expressions when the $N_1$-$N_2$ overlap effects were accounted for. In Sec.~\ref{sec:Gosc} we included the effects of the $N_1$-$N_2$ oscillation. We showed that the CP violation $A_{\rm CP} \propto (\Gamma_{-} - \Gamma_{+})$ is nonzero if the phase difference $\Delta \phi \not=0$ where $\Delta \phi = \phi_2 - \phi_1$ and $\phi_j$ is the phase of the heavy-light mixing parameter $U_{\ell_2 N_j}$, cf.~Eqs.~(\ref{delphi}), (\ref{simpl}), (\ref{Geffexpr}).

In numerical evaluations, we considered that the produced charged lepton is $\tau$ ($\ell = \tau$), therefore the heavy-light mixing parameter $|U_{\ell N}|^2$ has less restrictive upper bounds than in the case of $\ell=e, \mu$. Further, for the heavy-light Yukawa couplings ${\cal A}_j^{(k)}$ of the first vertex [Eq.~(\ref{Aj})] we assumed values $\sim 10^{-4}$ related by the corresponding estimate of Eq.~(\ref{Gnutauest}) of the neutrino Yukawa parameter $G_{\nu_{\tau}}$. We found out that under such conditions, the mentioned rare decays of Higgs and the corresponding CP-violating ratio $A_{\rm CP}$ can be detected at the future International Linear Collider (ILC), although with low number of events $\sim 10^1$.


\appendix
\section{Decay width $\Gamma(h \to \nu_k \ell_2 \pi)$ in a general case of two Majorana neutrinos}
\label{app:Gd1exact}

In this Appendix we present calculation of the decay width $\Gamma(h \to \nu_k \ell_2^{\mp} \pi^{\pm})$ in the more general case where the two intermediate Majorana neutrinos $N_j$ ($j=1,2$) are not necessarily almost degenerate. This means that we will not assume the relations (\ref{massdeg}). In particular, we will not assume that $\Delta M_N \lesssim \Gamma_{N_j}$. Nonetheless, we will consider that $\Gamma_{N_j} \ll M_{N_j}$.

The squares of the decay amplitudes (\ref{Tsq}) enter the integration (\ref{dGamma1}) over the phase space of the final particles. The integration over $d \Omega_{{\hat p}^{''}_2}$ affects in a nontrivial way only the factor $(p_1 \cdot p_2)$ in ${\overline T}(p_N^2; p_1 \cdot p_2)_{i j}$ Eq.~(\ref{Texpr})
\be
\int d \Omega_{{\hat p}^{''}_2} {\overline T}(p_N^2; p_1 \cdot p_2)_{i j}
= (4  \pi)  {\overline T}(p_N^2; p_1''^{0} p_2''^{0})_{i j},
\label{dOmbTij} \ee
where $ p_1''^{0} p_2''^{0}$ are the energies of $\nu_k$ and $\ell_2$ in the $N_j$ rest system ($\Sigma''$)
\bea
p_1''^{0} & = & \frac{1}{2 \sqrt{p_N^2}} (M_h^2 - p_N^2), \qquad
p_2''^{0} =  \frac{1}{2 \sqrt{p_N^2}} ( p_N^2 + m_{\ell_2}^2 - m_{\pi}^2 ).
\label{p1p20pp} \eea
The subsequent integration over $d \Omega_{{\hat p}^{'}_N}$ is trivial, i.e., it gives only the overall factor of $(4 \pi)$.

This then leads, after some algebra, to the following expression [${\overline T}(p_N^2, p_1 \cdot p_2)_{i j} \mapsto {\overline {\cal T}}(p_N^2)_{i j}$]:
\bes
\label{bcalT}
\bea
 {\overline {\cal T}}(p_N^2)_{i j} & \equiv & \frac{1}{ (4 \pi)^2 }
 \int d \Omega_{{\hat p}^{'}_N} \int d \Omega_{{\hat p}^{''}_2}  {\overline T}(p_N^2, p_1 \cdot p_2)_{i j}
   \label{bcalTdef} \\
   & = & \frac{1}{2} (p_N^2)^2 M_h^2  \left( 1 - \frac{p_N^2}{M_h^2} \right)
   \left[ 1 - y_{\pi} - 2 y_{\ell_2} - y_{\ell_2} (y_{\pi} - y_{\ell_2}) \right]
   \left[ 1 - \frac{(p_N^2 - M_{N_i} M_{N_j})}{2 p_N^2} \right],
\label{bcalTres} \eea \ees
where we denoted
\be
y_{\ell_2} = \frac{m_{\ell_2}^2}{p_N^2}, \quad y_{\pi} = \frac{m_{\pi}^2}{p_N^2}.
\label{ys} \ee
Using this expression, we obtain [using Eqs.~(\ref{Tsq}) with $\Gamma_W \ll M_W$, and Eq.~(\ref{dGamma1})] after some algebra
\bea
\lefteqn{
\frac{d \Gamma(h \to \nu_k \ell_2 \pi)}{d p_N^2} =
\frac{K^2}{64 \pi^2} \frac{(p_N^2)^2 M_h}{M_{N_1} \Gamma_N} \left( 1 - \frac{p_N^2}{M_h^2} \right)^2  \left[ 1 - y_{\pi} - 2 y_{\ell_2} - y_{\ell_2} (y_{\pi} - y_{\ell_2}) \right] \lambda^{1/2}(1, y_{\ell_2}, y_{\pi})
}
\nonumber \\ && \times
{\bigg \{} \frac{\Gamma_N}{\Gamma_{N_1}} ({\cal A}^{(k)}_1)^2  |U_{\ell_2 N_1}|^2 \delta( p_N^2 - M_{N_1}^2) \left( 1 - \frac{(p_N^2 - M_{N_1}^2)}{2 p_N^2} \right)
\nonumber\\ &&
+ \frac{\Gamma_N M_{N_1} }{\Gamma_{N_2} M_{N_2}}  ({\cal A}^{(k)}_2)^2 |U_{\ell_2 N_2}|^2\delta( p_N^2 - M_{N_2}^2) \left( 1 - \frac{(p_N^2 - M_{N_2}^2)}{2 p_N^2} \right)
\nonumber\\ &&
+
2 {\cal A}^{(k)}_1 {\cal A}^{(k)}_2 |U_{\ell_2 N_1}| |U_{\ell_2 N_2}|
\left[ \frac{M_{N_1} \Gamma_N}{\pi} {\rm Re}(P_1 P_2^{\ast}) \cos(\Delta \phi) \pm \frac{M_{N_1} \Gamma_N}{\pi} {\rm Im}(P_1 P_2^{\ast})  \sin ( \Delta \phi) \right] \left( 1 - \frac{(p_N^2 - M_{N_1} M_{N_2})}{2 p_N^2} \right) {\bigg \}},
\nonumber\\
\label{dGd1exact}
\eea
where one should keep in mind that $y_{\ell_2}$ and $y_{\pi}$ are $p_N^2$-dependent, cf.~Eq.~(\ref{ys}). In the terms containing $P_1 P_1^{\ast}$ and $P_2 P_2^{\ast}$ we took into account
\be
P_j P_j^{\ast} = \frac{\pi}{ M_{N_j} \Gamma_{N_j} } \delta( p_N^2 - M_{N_j}^2) \quad (j=1,2),
\label{PjPjgen} \ee
because $\Gamma_{N_j} \ll M_{N_j}$ [cf.~also Eq.~(\ref{PjPj})].

In the expression (\ref{dGd1exact}) we see that in the terms containing $P_j P_j^{\ast}$ ($j=1,2$) the terms $\propto (p_N^2 - M_{N_j})$ give zero, i.e.,
\be
 \delta (p_N^2 - M_{N_j}^2) \left( 1 - \frac{(p_N^2 - M_{N_j}^2)}{2 p_N^2} \right)
 = \delta (p_N^2 - M_{N_j}^2) \quad (j=1,2).
\label{deltatozero} \ee
On the other hand, in the terms in Eq.~(\ref{dGd1exact}) that contain $P_1 P_2^{\ast}$ (i.e., the overlap terms), we cannot a priori claim that the terms $\propto (p_N^2 - M_{N_1} M_{N_2})$ give zero. Only in the case of near degeneracy $0 < \Delta M_N \lesssim \Gamma_N$, Eq.~(\ref{massdeg}), where the formulas (\ref{ImP1P2})-(\ref{ReP1P2}) are valid, the terms  $\propto (p_N^2 - M_{N_1} M_{N_2})$ give zero.

When the integration over $p_N^2$ is performed
\bea
\label{Gd1exact}
\Gamma(h \to \nu_k \ell_2 \pi) & = & \int_{(m_{\ell_2}+m_{\pi})^2}^{M_h^2} d p_N^2 \; \frac{d \Gamma(h \to \nu_k \ell_2 \pi)}{d p_N^2},
\eea
then in the terms with $P_1 P_1^{\ast}$ and $P_2 P_2^{\ast}$ of Eq.~(\ref{dGd1exact}) the Dirac-delta disappears and $p_N^2$ is replaced by $M_{N_1}^2$ and $M_{N_2}^2$, respectively. In the overlap term, i.e. the term containing $P_1 P_2^{\ast}$,  in general the integration over $d p_N^2$ must be performed numerically; however, in the case of near degeneracy, $0 < \Delta M_N \lesssim \Gamma_N$, Eqs.~(\ref{eta})-(\ref{del}) can be used, and the expression (\ref{Gd1exact}) reduces directly to the expression (\ref{Gd1}).

\section{The overlap function $\delta(y)$}
\label{app:delta}

The numerical results for the overlap functions $\eta(y)/y$ and $\delta(y)$, defined via Eqs.~(\ref{PP}) and under the conditions (\ref{massdeg}), were obtained numerically in the works \cite{GCCSKJZS1,GCCSKJZS2}\footnote{In Ref.~\cite{GCCSKJZS1} rare decays of $\pi^{\pm}$, and in Ref.~\cite{GCCSKJZS2} rare decays of heavier pseudoscalar mesons were considered, in both cases in the scenario of two almost degenerate on-shell neutrinos.}
and can be shown to agree with the explicit expressions (\ref{yetdel}) to a high precision (of $\sim 1\%$). The explicit expression (\ref{eta}) for $\eta(y)/y$ was derived in \cite{symm} (Appendix A.6 there). In \cite{CPVBelle} it was noted (and used) that the numerical results of Refs.~\cite{GCCSKJZS1,GCCSKJZS2,symm} for the overlap function $\delta(y)$ can be fitted with the expression (\ref{del}) to a high precision.

Here we derive the explicit expression (\ref{del}) for $\delta(y)$, under the assumption of the almost mass degenerate scenario (\ref{massdeg}). As this function is defined via Eq.~(\ref{ReP1P2}) [with notation of Eq.~(\ref{Pj})] involving ${\rm Re}(P_1 P_2^{\ast})$, explicit evaluation gives for this expression a sum of two terms
\be
{\rm Re}(P_1 P_2^{\ast}) =  R_1 + R_2,
\label{R1R2sum} \ee
where
\bes
\label{R1R2}
\bea
R_1 & \equiv & \frac{ \Gamma_{N_1} \Gamma_{N_2} M_{N_1} M_{N_2} }{ \left[ (p_{N}^2 - M_{N_1}^2)^2 + \Gamma_{N_1}^2 M_{N_1}^2 \right] \left[ (p_{N}^2 - M_{N_2}^2)^2 + \Gamma_{N_2}^2 M_{N_2}^2 \right]}
\nonumber\\
  &=& \frac{ \epsilon^2 \xi (2 - \xi) (1 + \epsilon y) M_N^4 }{ \left[ \left( x - M_N^2 \right)^2 + \epsilon^2 \xi^2 M_N^4 \right] \left[ \left( x - M_N^2 (1 + \epsilon y)^2 \right)^2 + \epsilon^2 (2 - \xi)^2 (1 + \epsilon y)^2 M_N^4 \right]},
\label{R1} \\
R_2 & \equiv & \frac{ (p_N^2 - M_{N_1}^2)  (p_N^2 - M_{N_2}^2) }{ \left[ (p_{N}^2 - M_{N_1}^2)^2 + \Gamma_{N_1}^2 M_{N_1}^2 \right] \left[ (p_{N}^2 - M_{N_2}^2)^2 + \Gamma_{N_2}^2 M_{N_2}^2 \right]}
\nonumber\\
  &=& \frac{ (x - M_N^2) \left( x - M_N^2 (1 + \epsilon y)^2 \right) }{ \left[ \left( x - M_N^2 \right)^2 + \epsilon^2 \xi^2 M_N^4 \right] \left[ \left( x - M_N^2 (1 + \epsilon y)^2 \right)^2 + \epsilon^2 (2 - \xi)^2 (1 + \epsilon y)^2 M_N^4 \right]},
\label{R2}
\eea \ees
where we used in Eqs.~(\ref{R1})-(\ref{R2}) the quasidegeneracy assumption (\ref{deg}), the notations (\ref{massdeg}), (\ref{yGN}) and
\bes
\label{notapp}
\bea
x & \equiv & p_N^2, \quad \Gamma_N = \epsilon M_N,
\label{notappa} \\
\Gamma_{N_1} &=& \xi \Gamma_N = \epsilon \xi M_N, \qquad
\Gamma_{N_2} = (2 - \xi) \Gamma_N = \epsilon (2 - \xi) M_N  \qquad (0 < \xi < 2)
\label{notappb} \\
\Rightarrow (\Delta M_N) &=& y \Gamma_N = \epsilon y M_N, \quad M_{N_1}=M_N (1 + \epsilon y).
\label{notappc} \eea \ees
The quasidegeneracy ascenario (\ref{massdeg}) thus means that $\epsilon \ll 1$ (and $y \lesssim 1$).
The explicit integration of $R_1$ over $x$ then gives
\bes
\label{I1}
\bea
I_1  &\equiv&  \int_{-\infty}^{+\infty} d x R_1(x)
=   \frac{\pi}{2 \epsilon M_N^2} \frac{\left[ 1 + \frac{1}{2} \epsilon y (2 - \xi) \right]}{ \left\{ (1 + y^2) + \epsilon y (2 - \xi + y^2) + \frac{1}{4} \epsilon^2 y^2 \left[ (2 - \xi)^2 + y^2 \right] \right\} }
\label{I1exact} \\
&=& \frac{\pi}{2 M_N \Gamma_N} \frac{\left[ 1 + {\cal O}(\epsilon) \right]}{
    \left[ 1 + y^2 + {\cal O}(\epsilon) \right]} = \frac{\pi}{2 M_N \Gamma_N} \frac{1}{(1 + y^2)} \qquad (\epsilon \ll 1).
\label{I1expr} \eea \ees

It turns out that the integration of $R_2$ over $x$ gives exactly the same result [including the $\epsilon$-dependent corrections in Eq.~(\ref{I1exact})]
\be
I_2  \equiv  \int_{-\infty}^{+\infty} d x R_2(x) =   I_1 =  \frac{\pi}{2 M_N \Gamma_N} \frac{1}{(1 + y^2)}.
\label{I2expr} \ee

Furthermore, it can be checked that in both cases, the integrands $R_1(x)$ and $R_2(x)$ give the most dominant contribution around $x \approx M_N^2$, in an interval around $M_N^2$ of length $\sim \kappa M_N \Gamma_N$. This is so because the tails (outside such an interval) give
\bes
\label{tails}
\bea
{\rm tail} \; R_1 & \sim & \int_{M_N^2 + \kappa M_N \Gamma_N}^{+\infty} dx \frac{\Gamma_N^2 M_N^2}{(x - M_N^2)^4} \sim \frac{1}{\kappa^3} \frac{1}{M_N \Gamma_N},
\label{tI1} \\
{\rm tail} \; R_2 & \sim & \int_{M_N^2 + \kappa M_N \Gamma_N}^{+\infty} dx \frac{1}{(x - M_N^2)^2} \sim \frac{1}{\kappa} \frac{1}{M_N \Gamma_N}.
\label{tI2}
\eea \ees
When $\kappa$ increases (e.g., $\kappa \sim 10$), these tail contributions are small in comparison to the full contributions (\ref{I1expr}) and (\ref{I2expr}) (we recall that $y \lesssim 1$). However, the tail contribution of $R_2$, although small ($\sim 1/\kappa$) , is significantly larger than the tail contribution of $R_1$ ($\sim 1/\kappa^3$).

Equations (\ref{I1expr}) and (\ref{I2expr}), together with the tail suppression (\ref{tails}), then imply the final result
\be
 {\rm Re}(P_1 P_2^{\ast}) \approx \frac{1}{(1 + y^2)} \frac{\pi}{M_N \Gamma_N} \delta(p_N^2 - M_N^2),
\label{ReP1P2fin} \ee
i.e., Eqs.~(\ref{ReP1P2}) and (\ref{del}).

In this Appendix we integrated $x$ ($\equiv p_N^2$) over the entire real axis, although the kinematic restrictions in the considered rare decays are $(0 \approx) (m_{\ell_2} + m_{\pi})^2 \leq p_N^2 \leq M_h^2$. The bound $M_h^2 \sim 10^4 \ {\rm GeV}^2$ is usually very far above $M_N^2$, and the lower bound very far below $M_N^2$ (in units of $\Gamma_N M_N$).

The other more complex problem is that the expression ${\rm Re}(P_1 P_2^{\ast})$ is in general not multiplied by a $p_N^2$-independent constant when integrated over $d p_N^2$ during the integration over the final particle phase space [cf.~Eq.~(\ref{d3})]. For example, in the problem considered in this work, it is multiplied by a $p_N^2$-dependent trace Eq.~(\ref{Texpr}), or more precisely, by the quantity Eq.~(\ref{bcalT}). In general, the less suppressed tail contribution (\ref{tI2}) may then become enhanced and may affect the results, i.e.,  the approximation (\ref{ReP1P2fin}) would need some nonnegligible corrections. High precision numerical calculations involving the phase space integral\footnote{
See Eqs.~(\ref{bcalT})-(\ref{dGd1exact}).}
\[
\int_{(m_{\ell_2}+m_{\pi}^2}^{M_h^2} d p_n^2 {\overline {\cal T}}(p_N^2)_{1 2} {\rm Re}(P_1 P_2^{\ast}),
\]
where for $ {\rm Re}(P_1 P_2^{\ast})$ we use the explicit expression Eqs.~(\ref{R1R2sum})-(\ref{R1R2}),
indicate that the approximation (\ref{ReP1P2fin}) is very good for $M_N \gtrsim 1$ GeV (keeping in mind that $M_h \approx 125$ GeV). If we require that the formula (\ref{ReP1P2fin}) reproduce the correct result with an error less than 1 \%, we need to have $\epsilon (\equiv \Gamma_N/M_N) \sim 10^{-4}$ when $M_N \gtrsim 10$ GeV, and $\epsilon \sim 10^{-6}$ when $M_N \sim 1$ GeV.

\newpage
\begin{acknowledgments}
\noindent
This work was supported in part by FONDECYT Grants No.~1180344 (G.C.). The work of J.Z-S. was funded by ANID - Millennium Program - ICN2019\_044. The work of CSK is supported by NRF of Korea (NRF-2021R1A4A2001897).
\end{acknowledgments}

\end{document}